\documentclass[twocolumn,times]{aastex62}
\usepackage{url}
\usepackage{enumitem}
\usepackage{amsmath,amssymb}
\usepackage{booktabs}
\usepackage{textcomp}
\usepackage{cleveref}
\hypersetup{breaklinks,colorlinks,citecolor=blue,linkcolor=blue}

\newcommand{\kms}{\ensuremath{\rm{km}\, \rm{s}^{-1}}}

\newcommand{\Msun}{\ensuremath{\rm{M}_{\odot}}}
\newcommand{\Mstar}{\ensuremath{\rm{M}_{*}}}

\newcommand{\aco}{\alpha_\mathrm{CO}}

\newcommand\as{{\ooalign{{$.$}\cr{$''$}}}}

\usepackage[caption=false]{subfig}

\usepackage[nolist, nohyperlinks]{acronym}

\accepted{July 6, 2020}
\submitjournal{ApJ}
\shorttitle{Comparing the Stellar, CO and Dust-Continuum Emission of $z\sim 2$ Galaxies}
\shortauthors{Kaasinen et al.}

\graphicspath{.}

\begin{document}

\begin{acronym}
\acro{ism}[ISM]{interstellar medium}
\acro{sed}[SED]{Spectral Energy Distribution}
\acro{sfr}[SFR]{star formation rate}
\acro{alma}[ALMA]{Atacama Large Millimeter/submillimeter Array}
\acro{noema}[NOEMA]{NOrthern Extended Millimeter Array}
\acro{aspecs}[ASPECS]{ALMA Spectroscopic Survey in the Hubble Ultra Deep Field}
\acro{hudf}[HUDF]{Hubble Ultra Deep Field}
\acro{casa}[{\sc{CASA}}]{Common Astronomy Software Application \citep{2007ASPC..376..127M}} 
\acro{smgs}[SMGs; \citealt{Blain_2002}]{sub-millimetre selected galaxies}
\acro{dsfgs}[DSFGs]{dusty star-forming galaxies \cite{2014PhR...541...45C}}
\acro{hst}[\emph{HST}]{\emph{Hubble Space Telescope}}
\acro{IR}[IR]{infrared}
\acro{FIR}[FIR]{far-infared}
\acro{UV}[UV]{ultraviolet}
\acro{rms}[rms]{root mean square}
\acro{sn}[S/N]{signal-to-noise ratio}
\end{acronym}

\title{A Comparison of the Stellar, CO and Dust-Continuum Emission from Three, Star-Forming HUDF Galaxies at $z\sim 2$}

\author{Melanie Kaasinen}
\affil{Max-Planck-Institut f\"{u}r Astronomie, K\"{o}nigstuhl 17, D-69117 Heidelberg, Germany}
\affil{Universit\"{a}t Heidelberg, Zentrum f\"{u}r Astronomie, Institut f\"{u}r Theoretische Astrophysik, Albert-Ueberle-Stra\ss e 2, D-69120 Heidelberg, Germany}
\affil{ARC Centre of Excellence for All Sky Astrophysics in 3 Dimensions (ASTRO 3D)}
\author{Fabian Walter}
\affil{Max-Planck-Institut f\"{u}r Astronomie, K\"{o}nigstuhl 17, D-69117 Heidelberg, Germany}
\author{Mladen Novak}
\affil{Max-Planck-Institut f\"{u}r Astronomie, K\"{o}nigstuhl 17, D-69117 Heidelberg, Germany}
\author{Marcel Neeleman}
\affil{Max-Planck-Institut f\"{u}r Astronomie, K\"{o}nigstuhl 17, D-69117 Heidelberg, Germany}
\author{Ian Smail}
\affil{Centre for Extragalactic Astronomy, Department of Physics, Durham University, Durham DH1 3LE, UK}
\author{Leindert Boogaard}
\affil{Leiden Observatory, Leiden University, P.O. Box 9513, NL-2300 RA Leiden}
\author{Elisabete da Cunha}
\affil{International Centre for Radio Astronomy Research, University of Western Australia, 35 Stirling Hwy, Crawley, WA 6009, Australia
}
\affil{Research School of Astronomy and Astrophysics, The Australian National University, Canberra, ACT 2611, Australia}
\affil{ARC Centre of Excellence for All Sky Astrophysics in 3 Dimensions (ASTRO 3D)}
\author{Axel Weiss}
\affil{Max-Planck-Institut f\"{u}r Radioastronomie, Auf dem H\"{u}gel 69, 53121 Bonn, Germany}
\author{Daizhong Liu}
\affil{Max-Planck-Institut f\"{u}r Astronomie, K\"{o}nigstuhl 17, D-69117 Heidelberg, Germany}
\author{Roberto Decarli}
\affil{INAF—Osservatorio di Astrofisica e Scienza dello Spazio di Bologna, via Gobetti 93/3, I-40129, Bologna, Italy}
\author{Gerg\"{o} Popping}
\affil{European Southern Observatory, Karl-Schwarzschild-Strasse 2, 85748, Garching, Germany}
\author{Tanio Diaz-Santos}
\affil{N\'{u}cleo de Astronom\'{i}a, Facultad de Ingenier\'{i}a y Ciencias, Universidad Diego Portales, Ej\'{e}rcito Libertador 441, Santiago 8320000,
Chile}
\affil{Institute of Astrophysics, Foundation for Research and Technology-Hellas (FORTH), Heraklion, GR-70013, Greece}
\affil{Chinese Academy of Sciences South America Center for Astronomy (CASSACA), National Astronomical Observatories, CAS, Beijing 100101, China}
\author{Paulo Cort\'{e}s}
\affil{Joint ALMA Observatory - ESO, Av. Alonso de C\'{o}rdova, 3107, Santiago, Chile}
\affil{National Radio Astronomy Observatory, 520 Edgemont Rd, Charlottesville, VA, 22903, USA}
\author{Manuel Aravena}
\affil{N\'{u}cleo de Astronom\'{i}a, Facultad de Ingenier\'{i}a y Ciencias, Universidad Diego Portales, Ej\'{e}rcito Libertador 441, Santiago 8320000,
Chile}
\author{Paul van der Werf}
\affil{Leiden Observatory, Leiden University, P.O. Box 9513, NL-2300 RA Leiden}
\author{Dominik Riechers}
\affil{Department of Astronomy, Cornell University, Space Sciences Building, Ithaca, NY 14853, USA}
\affil{Max-Planck-Institut f\"{u}r Astronomie, K\"{o}nigstuhl 17, D-69117 Heidelberg, Germany}
\author{Hanae Inami}
\affil{Hiroshima Astrophysical Science Center, Hiroshima University, 1-3-1 Kagamiyama, Higashi-Hiroshima, Hiroshima 739-8526, Japan}
\author{Jacqueline A. Hodge}
\affil{Leiden Observatory, Leiden University, P.O. Box 9513, NL-2300 RA Leiden}
\author{Hans-Walter Rix}
\affil{Max-Planck-Institut f\"{u}r Astronomie, K\"{o}nigstuhl 17, D-69117 Heidelberg, Germany}
\author{Pierre Cox}
\affil{Institut dastrophysique de Paris, Sorbonne Universit\'{e}, CNRS, UMR 7095, 98 bis bd Arago, 7014 Paris, France}




\begin{abstract}
	We compare the extent of the dust, molecular gas and stars in three star-forming galaxies, at $z= 1.4, 1.6$ and $2.7$, selected from the Hubble Ultra Deep Field based on their bright CO and dust-continuum emission as well as their large rest-frame optical sizes. The galaxies have high stellar masses, $\Mstar>10^{11}\Msun$, and reside on, or slightly below, the main sequence of star-forming galaxies at their respective redshifts. We probe the dust and molecular gas using subarcsecond Atacama Large Millimeter/submillimeter Array observations of the 1.3 mm continuum and CO line emission, respectively, and probe the stellar distribution using \emph{Hubble Space Telescope} observations at 1.6 \textmu m. We find that for all three galaxies the CO emission appears $\gtrsim 30\%$ more compact than the stellar emission. For the $z= 1.4$ and $2.7$ galaxies, the dust emission is also more compact, by $\gtrsim 50\%$, than the stellar emission, whereas for the $z=1.6$ galaxy, the dust and stellar emission have similar spatial extents. This similar spatial extent is consistent with observations of local disk galaxies. However, most high redshift observations show more compact dust emission, likely due to the ubiquity of central starbursts at high redshift and the limited sensitivity of many of these observations. Using the CO emission line, we also investigate the kinematics of the cold interstellar medium in the galaxies, and find that all three have kinematics consistent with a rotation-dominated disk.
\end{abstract}


\section{Introduction}
	\label{sec:intro}

	The molecular gas phase of the \ac{ism} is a crucial component of star-forming galaxies, hosting and providing the fuel for star formation. Constraining the spatial distribution of both stars and molecular gas is therefore critical to understanding the evolutionary state of a galaxy. 
	Whereas the stellar component of a galaxy is best traced via the rest-frame near-infrared emission, molecular gas is most commonly observed via carbon monoxide (CO) line emission or \ac{FIR} dust-continuum emission. These stellar and molecular gas tracers have been mapped at high resolution (down to 100 pc scales) for local galaxies \citep[e.g.][]{2008AJ....136.2782L,2013ApJ...777....5S,2019Msngr.177...36S}, providing fundamental insights into the physics of star formation and the matter cycle of the \ac{ism}. However, such detailed, multi-wavelength comparisons are still lacking for galaxies at the peak epoch of stellar mass assembly, at $z\sim 2$. 

	The advent of sub/millimeter interferometers, particularly the \ac{alma} and \ac{noema}, have led to a growing body of work aimed at characterising the molecular gas properties of $z\gtrsim 2$ galaxies via their CO and/or dust-continuum emission \citep[e.g. surveys described in][]{2016ApJ...833...67W,2017ApJ...837..150S,2018ApJ...853..179T,2018ApJ...864...49P,2019ApJ...882..138D,2019MNRAS.487.4648S}. Based on unresolved measurements, the long-wavelength (observed-frame 850 \textmu m), dust-continuum emission appears to trace the bulk of the cold, molecular gas in massive, star-forming galaxies as accurately as the traditional tracer, CO(1-0) \citep[][]{2014ApJ...783...84S,2019ApJ...880...15K}. However, resolved observations indicate that the dust-continuum emission stems from a significantly more compact region than the CO emission \citep[e.g][]{2015ApJ...799...81S,2015ApJ...798L..18H,2017ApJ...846..108C,2018ApJ...863...56C,2019MNRAS.490.4956G}, calling into question its application as a molecular gas tracer. 

	To date, the sizes of the rest-frame optical, dust-continuum and CO emitting regions have only been compared for the brightest and most massive high-redshift ($z>1$) sources. For $z\sim 2-4$ submillimeter-selected galaxies \citep[SMGs; ][]{2002PhR...369..111B}, the measured dust-continuum emission appears $2-4 \times$ more compact than the CO emission \citep{2015ApJ...799...81S,2015ApJ...798L..18H,2017ApJ...846..108C,2018ApJ...863...56C}. Similarly, for quasar host galaxies at $z\gtrsim 6$ (studied with lower resolution observations) the dust-continuum sizes also appear $z\sim 2-4$ more compact than those of the CO \citep{2018A&A...619A..39F,2019ApJ...880....2W}. The dust-continuum and rest-frame optical (but not CO) sizes, have also been compared in detail for SMGs \citep{2016ApJ...833..103H,2019ApJ...876..130H,2019MNRAS.490.4956G,2019ApJ...879...54L}, six, massive and compact star-forming galaxies at $z\sim 2.5$ \citep{2016ApJ...827L..32B},  a $z\sim 1.25$ starburst \citep{2019ApJ...870..130N} and massive, H$\mathrm{\alpha}$-selected galaxies at $z\sim 2.2-2.5$ \citep{2017ApJ...834..135T}. These studies all find that the dust emission is $2-4 \times$ more compact than the rest-frame optical emission.  

	It is still unclear how the spatial extents of the stellar, dust-continuum and CO emission compare for the wider population of star-forming galaxies, including the population conforming to the so-called ``main sequence'' (MS) of star-forming galaxies \citep[the correlation between the stellar mass and star formation rate of the majority of star-forming galaxies observed for each epoch up to $z\sim 4$, e.g. in][]{2007ApJ...660L..43N,2012ApJ...754L..14S,2012ApJ...754L..29W,2014ApJ...795..104W,2016A&A...589A..35S}. In the first study of MS galaxies, \cite{2017ApJ...841L..25T} measure $1.5-2\times$ smaller dust-continuum vs CO half-light radii. However, the two massive ($\Mstar>10^{11}\Msun$) galaxies in their study were selected based on their compact, dusty, star-forming cores. An additional complication to the selection biases, is that most previous observations of the dust-continuum emission of high redshift sources have been conducted at \ac{IR} wavelengths, at which the dust luminosity is highly sensitive to the ISM temperature and thereby traces the star formation rate along with the dust reservoir. Moreover, few studies exist with resolved  observations of both the CO and dust-continuum emission in unlensed objects \citep[e.g.][]{2015ApJ...798L..18H,2017ApJ...846..108C,2017ApJ...841L..25T,2018ApJ...863...56C}, of which only a subset are at a comparable resolution \citep{2015ApJ...798L..18H,2018ApJ...863...56C}. 

	In this work, we test how the spatial extent (quantified by the half-light radii) of dust continuum, CO and rest-frame optical emission compare for three extended, MS galaxies at $z\sim 2$. To this end, we selected sources in the \ac{hudf} for which the dust continuum and CO emission have already been detected with \ac{alma} as part of the \ac{aspecs} Large Program \citep[][]{2019ApJ...882..139G,2019ApJ...882..138D,2019ApJ...882..136A}. To best compare the spatial extents, we study the three sources with the most extended morphologies in the \ac{UV} to near \ac{IR} images taken with the \ac{hst}. We use observations taken as part of the \ac{aspecs} LP as well as our ``ALPS'' (ASPECS Large Program Source) follow-up, higher resolution \ac{alma} observations. Thus, throughout this paper we refer to our sources as ALPS.1, 2 and 3.

	This paper is structured as follows. In Section~\ref{sec:data} we describe the observations, data reduction and imaging of the CO and dust-continuum emission. We discuss the global galaxy properties inferred from the multi-wavelength data in Section~\ref{sec:derived_quantities}. In Section~\ref{sec:source_size_analysis}, we derive and compare the half-light radii of the dust-continuum emission, CO and rest-frame optical. We derive dynamical properties in Section~\ref{sec:dynamic_analysis} via kinematic modeling. We compare our sources to other samples in Section~\ref{sec:discussion} and summarise our main findings in Section~\ref{sec:conclusions}. Throughout this paper we assume a $\Lambda$CDM cosmology with $H_0 = 70\, \kms \mathrm{Mpc}^{-1}$, $\Omega_M = 0.3$ and $\Omega_\Lambda = 0.7$. Stellar masses and SFRs are based on a \cite{2003PASP..115..763C} IMF. When quoting molecular gas masses we account for a factor of 1.36 to include He.


	\begin{figure*}
		\centering
		\includegraphics[width=\textwidth, trim={1.2cm 0.5cm 1.3cm 1cm},clip]{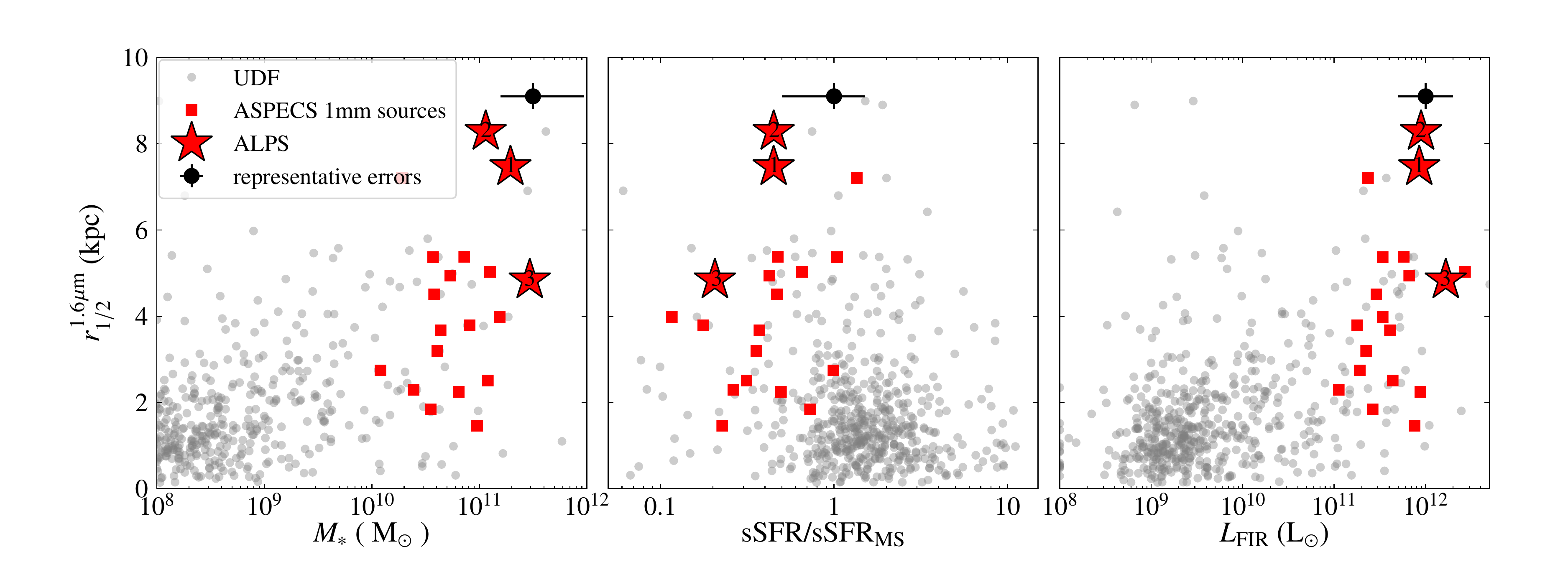}
		\caption{ Comparison of the ALPS sample properties relative to the parent samples at $1<z<3$. The 1.6\,\textmu m (stellar) half-light radii are shown as a function of stellar mass, offset from the main sequence and FIR luminosity (within 8-1000\textmu m) in the left, middle and right panels, respectively. The three ``ALPS'' sources studied in this work (red stars) are compared to the ASPECS 1 mm continuum and line sources (red squares) and the HUDF sample (grey circles), from \cite{2020arXiv200604284A}. The offset from the main sequence (in the middle panel) is calculated individually,  based on the measured redshift, stellar mass and SFR of each source, and, following \cite{2020arXiv200604284A}, the main sequence is defined by the best-fit relation of \cite{2015A&A...575A..74S}. The 1.6 \textmu m half-light radii are from Table 2 of \cite{2012ApJS..203...24V} and were measured from the F160W maps. The three galaxies studied here were selected to be both extended in the rest-frame optical and be CO- and FIR-bright. They also have large stellar masses and systematically lower sSFRs than the majority of galaxies observed in the HUDF. \label{fig:selection}}
	\end{figure*}

\newpage

\section{Sample Selection, Observation and Data Reduction} 
	\label{sec:data}

	\subsection{Galaxy Selection} 
		\label{sub:sample}

\begin{figure*}
		\centering
		\includegraphics[width=\textwidth, trim={1.7cm 2.cm 1.5cm 0.2cm},clip]{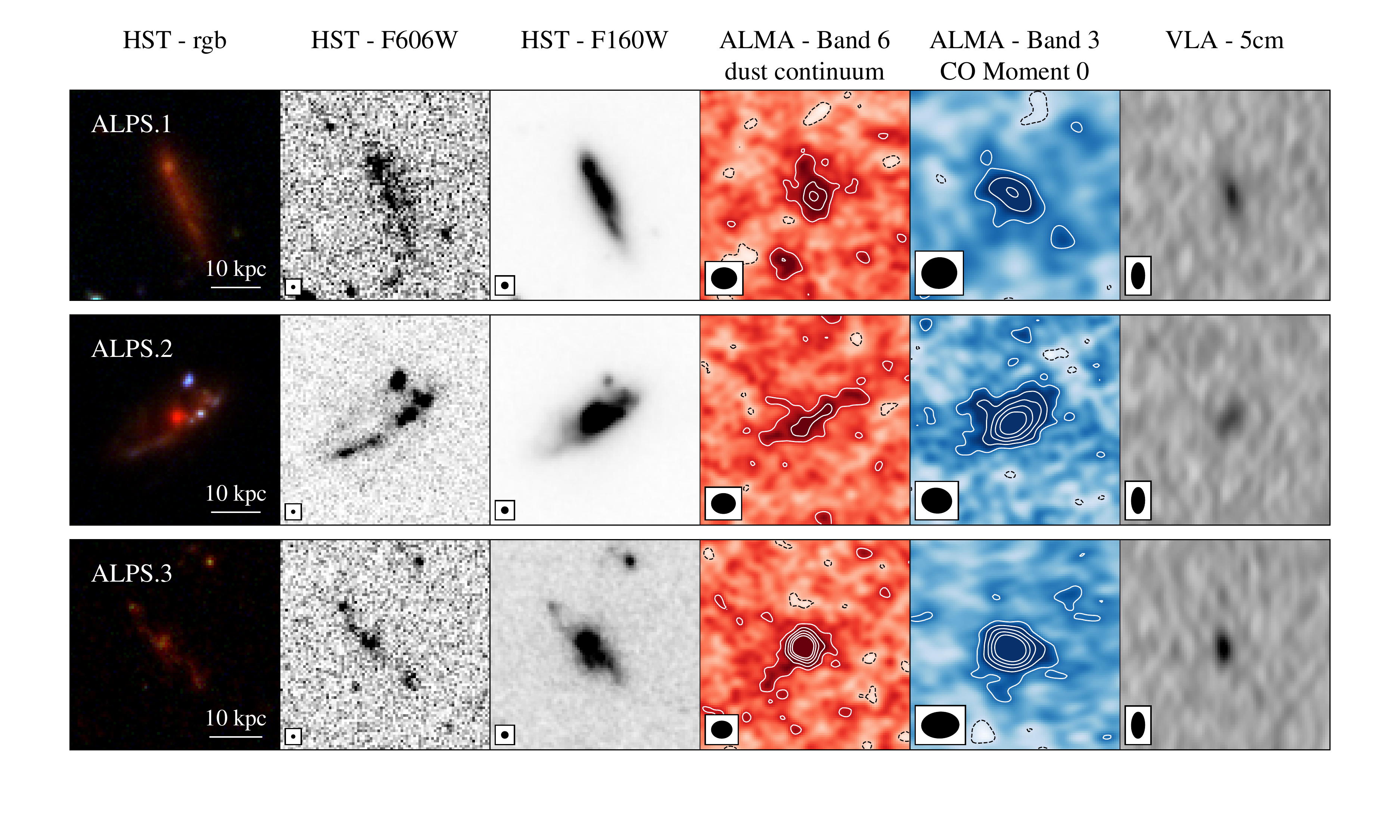}
		\caption{Multiwavelength data for the three observed sources (labeled ALPS 1, 2 and 3 at the upper left of the left panel in each row). Each panel depicts a $5" \times 5"$ region centred on the kinematic centre of the source. Columns, from left to right: HST 435-775-105 color composite, HST/F606W, HST/F160W, Band 6 (1.3\,mm) dust continuum (combined \ac{aspecs} LP and \citealt{2017MNRAS.466..861D} data), CO moment 0 map (from the combined ASPECS LP and ALPS data) and VLA - 5\,cm continuum flux. The contours for the Band 6 dust continuum and CO moment 0 start at $\pm 2\sigma$ and change in steps of $2\sigma$. Dashed black (solid white) lines show negative (positive) contours.  \label{fig:thumbnails}}
	\end{figure*}

		The three galaxies studied here are derived from the \ac{aspecs} Large Program (LP), a survey of the sub-mm to mm emission within the \ac{hudf} \citep{2019ApJ...882..139G,2019ApJ...882..138D}. Because the survey was conducted without preselection, it identified the sources brightest in CO line and dust-continuum emission. Based on the line searches performed by \cite{2019ApJ...882..139G}, ASPECS provided 16 highly significant CO detections. From this sample of 16, we selected the three CO- and dust-brightest galaxies, with the most extended rest-frame optical emission. These two criteria were equally important. We required the sources to be CO-bright in order to resolve their emission, at $\sim0\as5$, within five hours of integration time, with ALMA, and select galaxies that are extended in the rest-frame optical to increase the chances of observing extended dust and CO emission. In Figure \ref{fig:selection}, we compare the properties of the three galaxies studied here to the parent sample of ASPECS LP 1.2 mm candidates (listed in Table 1 of \citealt{2020arXiv200207199G}) and the full set of galaxies in the HUDF \citep[see][]{2020arXiv200604284A}. Our ``ALPS'' sources have large stellar masses ($\gtrsim 10^{11}\Msun$) and high FIR luminosities compared to the majority of the sources in the field, due to our selection of CO- and 1\,mm-bright targets. Their extended and inclined rest-frame UV-optical morphologies are shown in the first three columns of Figure \ref{fig:thumbnails}. 

		Compared to most observed SMGs, our galaxies have more extended morphologies in the rest-frame UV and optical emission. However, the interpolated 870\,\textmu m flux densities are consistent with the definition of an SMG, i.e. $>1$ mJy \citep[see][for discussions on the selection and classification]{2014PhR...541...45C,2020arXiv200400934H}. Assuming that the FIR continuum can be described by a modified blackbody with an average dust temperature of 25 K and a dust emissivity index of 1.8 (see Section \ref{sub:flux_densities}), the 1.3\,mm flux densities measured by \cite{2020arXiv200207199G} extrapolate to $1.4\pm 0.1$, $1.5\pm0.1$ and $2.8\pm 0.1$ mJy for ALPS.1, 2 and 3 respectively. However, the FIR luminosities (bolometric luminosity between 8-1000 \textmu m) of ALPS.1 and 2 are lower than what is typically measured for SMGs. Based on the models fit to the spectral energy distributions, ALPS.1, 2 and 3 have FIR luminosities of $(0.78\, \pm\, 0.04) \times 10^{12}$,  $(0.9\,\pm\, 0.1) \times 10^{12}$ and $(1.8\, \pm\, 0.6) \times 10^{12} \mathrm{L}_\odot$, respectively. SMGs have FIR luminosities of $L_\mathrm{FIR} \gtrsim 10^{12} \mathrm{L}_\odot$, typically $\sim 3 \times 10^{12} \mathrm{L}_\odot$ \citep[e.g.][]{2014MNRAS.438.1267S}. Thus, ALPS.1 and 2 are at the faint end of the SMG population, whereas ALPS.3 is entirely consistent with most observed SMGs. 
	

	\subsection{Dust continuum (Band 6) Observations} 
		\label{sub:Band_6_obs}

		We analyse two Band 6 (211-275 GHz; 1.1-1.4 mm) surveys of the HUDF, taken at different depths and resolutions. We use the unresolved observations from the ASPECS LP \citep[project code: 2016.1.00324, described in ][]{2020arXiv200207199G} and the 1.3\,mm observations of \cite{2017MNRAS.466..861D} (project code: 2012.1.00173.S), which provide a higher spatial resolution. The source IDs from the ASPECS LP and HUDF data of \cite{2017MNRAS.466..861D} are provided in Table~\ref{tab:source_prop}. In the subsequent analysis (Section~\ref{sub:data_reduction_and_imaging}), we combine both datasets to achieve high resolution without the loss of large-scale emission. The ASPECS 1.3\,mm observations, executed between March 10, 2017 and July 13, 2018, were conducted in the C43-3 configuration with a minimum baseline of 15 m and maximum baseline of 500 m. Our three sources lie towards the edges of the mosaic (where the primary beam reponses are 87\%, 65\% and 20\% of the peak sensitivity, respectively for ALPS.1, 2 and 3). Thus, the data for our sources are not as deep as those in the center of the mosaic. The \cite{2017MNRAS.466..861D} observations were conducted in \ac{alma} Cycle 2 using a variety of configurations, with maximum baselines between 550 and 1250 m \citep[see Table 2 of][]{2017MNRAS.466..861D}. The pointings of the ASPECS LP are more closely spaced than for the \cite{2017MNRAS.466..861D} observations, however the latter sample a larger area. Thus, for ALPS.3, on the edge of the ASPECS field, the majority of the data used here is from \cite{2017MNRAS.466..861D}.




	\begin{table} 
\begin{center} 
\caption{Description of ALMA Observations \label{tab:source_prop}} 
\begin{tabular}{@{}>{\raggedright}m{0.34\columnwidth}>{\centering}m{0.18\columnwidth}>{\centering}m{0.18\columnwidth}>{\centering\arraybackslash}m{0.18\columnwidth}@{}}
\toprule 
ALPS ID &ALPS.1 & ALPS.2 & ALPS.3\\ 
\midrule 
ASPECS LP 1\,mm ID &1mm.03 & 1mm.05 & 1mm.06\\ 
ASPECS LP 3\,mm ID &3mm.04 & 3mm.05 & 3mm.07\\ 
Dunlop et al. (2017) ID &UDF.6 & UDF.8 & UDF.2\\ 
\midrule 
RA (J2000) & 03:32:34.44 & 03:32:39.77 & 03:32:43.53\\ 
DEC (J2000) & -27:46:59.8 & -27:46:11.8 & -27:46:39.2\\ 
z$_\mathrm{CO}$  & $1.4140$ & $1.5504$ & $2.6961$\\ 
CO transition  & $2\rightarrow1$ & $2\rightarrow1$ & $3\rightarrow2$\\ 
\midrule 
\multicolumn{4}{@{}l}{50 km s$^{-1}$ CO Cube}\\ 
\midrule 
Weighting parameter  & $0.5$ & $2$ & $2$\\ 
Beam FWHM ($''\times''$)  & $0.82 \times 0.72$ & $0.71 \times 0.58$ & $0.88 \times 0.61$\\ 
rms per channel ~~~~ (mJy beam$^{-1}$) & 0.14 & 0.09 & 0.11\\ 
\midrule 
\multicolumn{4}{@{}l}{Dust continuum map (211-231 GHz; 1.3mm)}\\ 
\midrule 
Weighting parameter  & $2$ & $2$ & $2$\\ 
Beam FWHM ($''\times''$)  & $0.59 \times 0.49$ & $0.56 \times 0.47$ & $0.48 \times 0.40$\\ 
rms (\textmu Jy beam$^{-1}$) & 21 & 22 & 24\\ 
\bottomrule 
\end{tabular} 
\end{center} 
\end{table}

	\subsection{CO Line (Band 3) Observations} 
		\label{sub:Band_3_obs}

		To study the CO line emission of our three sources, we use two sets of ALMA Band 3 (84-115 GHz; 2.6-3.6 mm) data. We use the ASPECS LP Band 3 survey (project code: 2016.1.00324.L) and our targeted, follow-up ``ALPS'' observations, at higher resolution (project code: 2017.1.00270.S). The source IDs from the ASPECS LP are provided in Table~\ref{tab:source_prop} along with the CO-derived redshifts and observed CO transitions. An additional source was observed at the same redshift as, and at small angular separation to, ALPS.3 (ASPECS ID 3mm.09), but was excluded from the main text work due to the lack of extended emission. We discuss this source further in Appendix \ref{sec:additional_target}. 

		The ASPECS LP Band 3 observations comprised of 17 pointings conducted in the compact, C40-3 configuration (with baselines ranging from 15 to 700 m). Further details are described in \cite{2019ApJ...882..138D}. For our sample, the sensitivity of the observations was the highest for ALPS.2 and lowest for ALPS.3 (with primary beam responses of 70\%, 80\% and 50\% of the peak sensitivity for ALPS.1, 2 and 3, respectively).  In addition, we rely on new, supplementary, high-resolution Band 3 data, taken between January 4-11, 2018. These \ac{alma} observations were conducted in the C43-5 configuration (minimum baseline of 15 m and maximum baseline of 2.5 km). The Bandpass, flux and phase calibrators were J0522-3627, J0329-2357 and J0342-3007, respectively.


	\subsection{Data Reduction and Imaging} 
		\label{sub:data_reduction_and_imaging}

		All sets of raw data were reduced using the standard ALMA calibration scripts for the \ac{casa}. For all sets of observations, the standard pipeline produced $uv$-data products of high quality and was therefore used without additional modifications. Following the raw data reduction, we combine the various sets of data for each band. The subsequent concatenating and imaging was carried out using \ac{casa} version 5.4.0.

		\subsubsection{Dust Continuum (Band 6) Imaging} 
			\label{subsub:Band_6_imaging}

			We image the dust continuum using the Band 6 data from the \ac{aspecs} LP \citep{2020arXiv200207199G} and from \cite{2017MNRAS.466..861D}. These two datasets cover different frequency ranges, i.e. the \ac{aspecs} LP data \citep{2020arXiv200207199G} spans 212-272 GHz whereas the \cite{2017MNRAS.466..861D} data spans 211-231 GHz, and were taken with different configurations. In order to probe the extended 1.3\,mm emission at high resolution, and ensure that we are probing the same wavelengths at all radii, we select the four (out of 32) spectral windows from the ASPECS LP that match those of \cite{2017MNRAS.466..861D}. By subselecting these spectral windows from the ASPECS LP data, we reduce the depth of our observations relative to the LP. However, this selection is the best compromise between the sensitivity to extended emission and resolution required for our study. In effect this subselection is a form of tapering, which we use to avoid weighting the final maps to the large amount of short baseline measurements from the large programme.

			We jointly image the two sets of $uv$-data via \ac{casa}'s \texttt{TCLEAN}, by applying the ``mosaic'' gridding option. To maximise the \ac{sn} of the final images, we image the data using a natural weighting scheme and clean to $2\sigma$, within a circular mask of $3''$ radius. The cleaned maps, for each source, are shown in the 4th column of Figure \ref{fig:thumbnails} and their properties are summarised in Table~\ref{tab:source_prop}. The comparative size of the synthesized beams reflects the relative contribution of the ASPECS LP versus \cite{2017MNRAS.466..861D} data, with the beam size decreasing the further a source lies towards the edge of the ASPECS 1 mm mosaic. ALPS.3 (at the very edge of the LP mosaic) has by far the smallest beam size as most of the visibilities are from the \cite{2017MNRAS.466..861D} data.


		\subsubsection{CO Line (Band 3) Imaging} 
			\label{subsub:Band_3_co_imaging}

			To image the CO emission, we combine the low- and high-resolution data from the ASPECS LP Band 3 and ALPS programs.  We initially attempt to subtract the continuum (via \ac{casa}'s \texttt{uvcontsub}), but recover no significant continuum emission, rendering this step redundant. We image the $uv$-data using \ac{casa}'s \texttt{TCLEAN} task, mosaicing the pointings from the ASPECS LP and follow-up ALPS program via the ``mosaic'' gridding option and setting the phase centre to the expected source centre (the pointing of the ALPS data). We apply a circular mask of $3''$ radius to create all cleaned images. We test a variety of robust weighting parameters to find the optimum balance between the resolution and resulting noise level. For the final images of ALPS.2 and 3 we use natural weighting, whereas for ALPS.1 we use robust weighting (see Table~\ref{tab:source_prop}).

			To create the moment-0 maps, from which we estimate source sizes, we create a single-channel image over the extent of the line emission, (the 99.99th percentile range, as in \citealt{2019ApJ...882..136A}) applying the ``multiscale'' deconvolver (with a deconvolution scale of up to $2''$) and cleaning down to a threshold of $2 \sigma$ within a mask of radius $3''$. We compare these moment-0 maps to maps obtained from 50 and 100 $\kms$ data cubes, ensuring that the extent and morphology of the source emission in all moment-0 maps are consistent, but select the single-channel image for the moment-0 map, to ensure that we have robust noise estimates. 

	
		

\newpage

\section{Derived Global Quantities} 
	\label{sec:derived_quantities}

	\subsection{Deriving Dust and Gas Masses from the ALMA Maps} 
		\label{sub:flux_densities}

		We measure the total CO and dust-continuum flux densities from the cleaned, dirty and residual maps by applying the residual scaling method \citep[described in detail in Appendix A of][]{2019ApJ...881...63N}. This involves estimating the unknown, intrinsic flux by rescaling the contribution from the residual map with an estimate of the area of the dirty beam in the region of interest. The rescaling process is necessary to account for the ill-defined beam units of each cleaned interferometric map, which is a combination of the residual component in units of Jy per dirty beam, and a cleaned component in units of Jy per clean beam. For both the CO and dust-continuum emission, we extract the total flux densities from within circular apertures of $1\as5,~ 2''$ and $1\as5$ for ALPS.1, 2 and 3, respectively. We estimate the uncertainties of these flux densities as the local \ac{rms}, $\sigma$, (in units of Jy beam$^{-1}$) scaled by the square root of the number of independent clean beams filling the aperture. 

		We compare the flux densities measured here (Table \ref{tab:derived_prop}) to those derived using the ASPECS LP data only. Our CO emission line fluxes are consistent, within the uncertainties, with the spectral line fits of the unresolved \ac{aspecs} LP data, presented in \cite{Boogaard2020}, and are 20\%\ lower, on average, than the values in \cite{2019ApJ...882..139G} (although still consistent within errors). For ALPS.1 and 3, our 1.3\,mm continuum flux density measurements are consistent with the measurements of \cite{2017MNRAS.466..861D}, whereas our value for ALPS.2 is $(50\pm20)$\% larger. Conversely, the 1.3\,mm flux density measured here for ALPS.1 is $(55\pm 20)\%$ lower than the continuum flux density meaured based on the ASPECS LP data in \cite{2020arXiv200207199G}, whereas the flux density measured for ALPS.3 is consistent within the errors. The 1.3\,mm flux density measured here for ALPS.2 is consistent with the value in \cite{2020arXiv200207199G}. The difference between the flux densities measured by \cite{2017MNRAS.466..861D} and \cite{2020arXiv200207199G} can be mostly attributed to the different spectral setups, with the \cite{2017MNRAS.466..861D} covering longer wavelengths. Similarly, the smaller flux density measured here for ALPS.1 is, in part, the result of the subselection of spectral setups. 

		To estimate the molecular gas masses from our CO observations we first convert the line fluxes to luminosities (in $\mathrm{K}\, \kms \mathrm{pc}^{2}$) following \cite{1992ApJ...398L..29S}, via,
		\begin{align}
			L_\mathrm{CO}^{\prime} = 3.25 \times 10^7 S_\mathrm{CO} \Delta v \, \nu_\mathrm{obs}^{-2}  D_\mathrm{L}^2  (1+z)^{-3} \, ,
		\end{align} 
		where $S_\mathrm{CO} \Delta v$ is the velocity-integrated line flux (in Jy \kms), $\nu_\mathrm{obs}$ is the observed-frame frequency of the CO transition in GHz and $D_\mathrm{L}$ is the luminosity distance in Mpc. We down-convert these line luminosities to the CO(1-0) line luminosity via the following ratios (Table \ref{tab:derived_prop}). For ALPS.3, both the CO(3-2) and CO(1-0) transitions have been observed \citep{Riechers_2020}. Thus, we apply the measured luminosity ratio $r_\mathrm{31} = L_\mathrm{CO (3-2)}^{\prime}/L_\mathrm{CO(1-0)}^{\prime} = 0.79 \pm 0.21$. For ALPS.1 and 2 we have no CO(1-0) observations. Thus, we assume an excitation correction based on the measurements of $r_\mathrm{21} = L_\mathrm{CO (2-1)}^{\prime}/L_\mathrm{CO(1-0)}^{\prime}$ for other high-redshift samples \citep{2015A&A...577A..46D,2013MNRAS.429.3047B,2014ApJ...785..149S}, and, the values inferred for the ASPECS sample \citep{Boogaard2020}. 

		To convert the inferred CO(1-0) line luminosities to molecular gas masses we apply CO-to-molecular gas mass conversion factors, $\aco$, calculated via the assumed metallicities (provided in Table \ref{tab:derived_prop}). 
		We infer the metallicities (according to the \citealt{2004MNRAS.348L..59P} scale) from the stellar masses using the mass-metallicity calibration in Equation (12a) of \cite{2015ApJ...800...20G} (taking into account both the uncertainties on the measured stellar mass and the empirical relation). We note that for ALPS.2 the strong emission lines H$\alpha$ and [\ion{N}{2}] have been observed, yielding a slightly higher inferred metallicity than the value based on the mass-metallicity calibration \citep{2014MNRAS.443.3780W}. However, the difference is only $0.05$ dex, which is lower than the typical systematic uncertainty of metallicity measurements \citep[e.g.][]{2008ApJ...681.1183K}. We calculate the CO-to-molecular gas mass conversion factor, $\aco$, according to the Equation (2) of \cite{2018ApJ...853..179T}. We choose these mass-metallity and $\aco$ calibrations based on the detailed comparison of such relations presented in Appendix A.3 of \cite{2019ApJ...887..235L}.  The final molecular gas masses quoted in Table \ref{tab:derived_prop} are the inferred $L_\mathrm{CO(1-0)}^{\prime}$ multiplied by the metallicity-based $\aco$.

		We estimate the total dust masses in two ways; 1) from the SED fits (see Section \ref{sub:stellar_masses}) and 2) from the 1.3 mm (Band 6) flux densities. To convert the flux densities measured at 1.3 mm to dust masses we assume that the dust emission is optically thin and can be modeled by a modified blackbody of the form, 
		\begin{align}
			S_\mathrm{{\nu}_\mathrm{obs}} =  (1+z) D_L^{-2} \kappa_{{\nu}_\mathrm{rest}} M_\mathrm{dust} B_{{\nu}_\mathrm{rest}} (T_{\mathrm{dust}}) 
		\end{align} 
		where $\kappa_{{\nu}_\mathrm{rest}}$ is the dust mass opacity coefficient, $M_\mathrm{dust}$ is the dust mass, $B_{{\nu}_\mathrm{rest}}$ is the blackbody radiation spectrum and $T_{\mathrm{dust}}$ is the temperature of the dust.
		We assume a temperature of 25 K, as motivated in \cite{2014ApJ...783...84S}. We determine the opacity coefficient by relating it to a reference frequency via the power law dependency,
		\begin{align}
			\kappa_{{\nu}_\mathrm{rest}} = \kappa_{{\nu}_\mathrm{0}} (\nu_\mathrm{rest}/\nu_\mathrm{0}) ^ \beta , 
		\end{align}
		where $\beta$ is the dust spectral emissivity index. We use the rest-frame 850 \textmu m as our reference frequency, taking $\kappa_{{\nu}_{850 \mathrm{\mu m}}} = 0.77\, \mathrm{g^{-1}cm^2}$ \citep{1984ApJ...285...89D} and $\beta=1.8$ \citep[e.g.][]{2011A&A...536A..21P}. 

		For ALPS.1 and 2, the dust masses derived from the 1.3 mm flux densities are larger than the values inferred from the SED fitting (Table \ref{tab:derived_prop}), although still consistent within errors for ALPS.1. The discrepancy is the result of a combination of the assumed (or fit) temperatures and emissivity indices, which are affected by the AGN template used for ALPS.2. Unlike the single MBB model, {\sc{MAGPHYS}} includes both a warm and cold dust component with emissivity indices of 1.5 and 2 respectively. The temperature of the two components is determined during the fit. Thus, the mass-weighted temperature is not fixed. For ALPS.2, in particular, the mass-weighted temperature of the {\sc{MAGPHYS}} fit is greater than for the single MBB, resulting in the difference between the two values.


		\begin{figure}
			\centering
			\includegraphics[width=0.48\textwidth, trim={0.1cm 0.cm 0.5cm 0.1cm},clip]{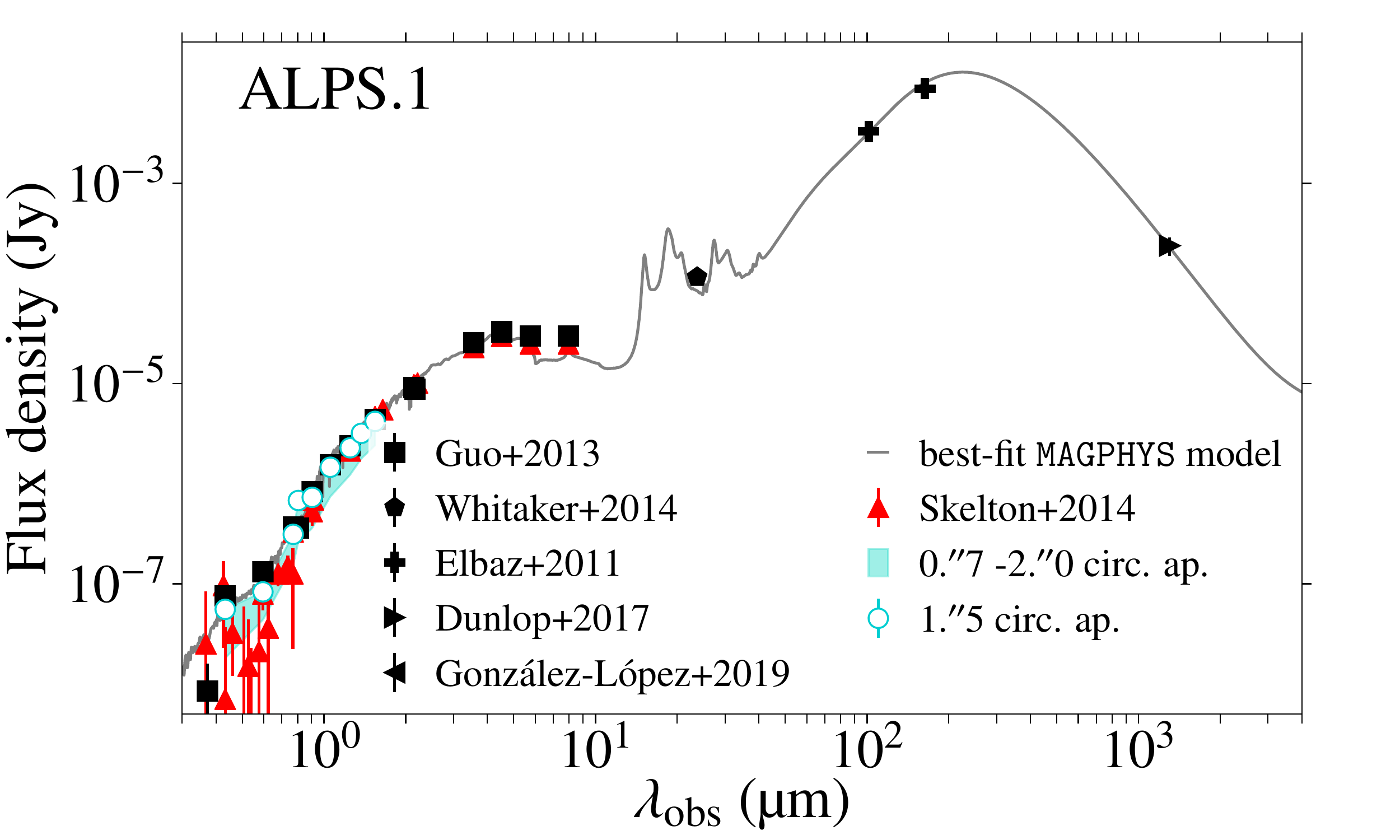}
			\\
			\includegraphics[width=0.48\textwidth, trim={0.1cm 0.cm 0.5cm 0.1cm},clip]{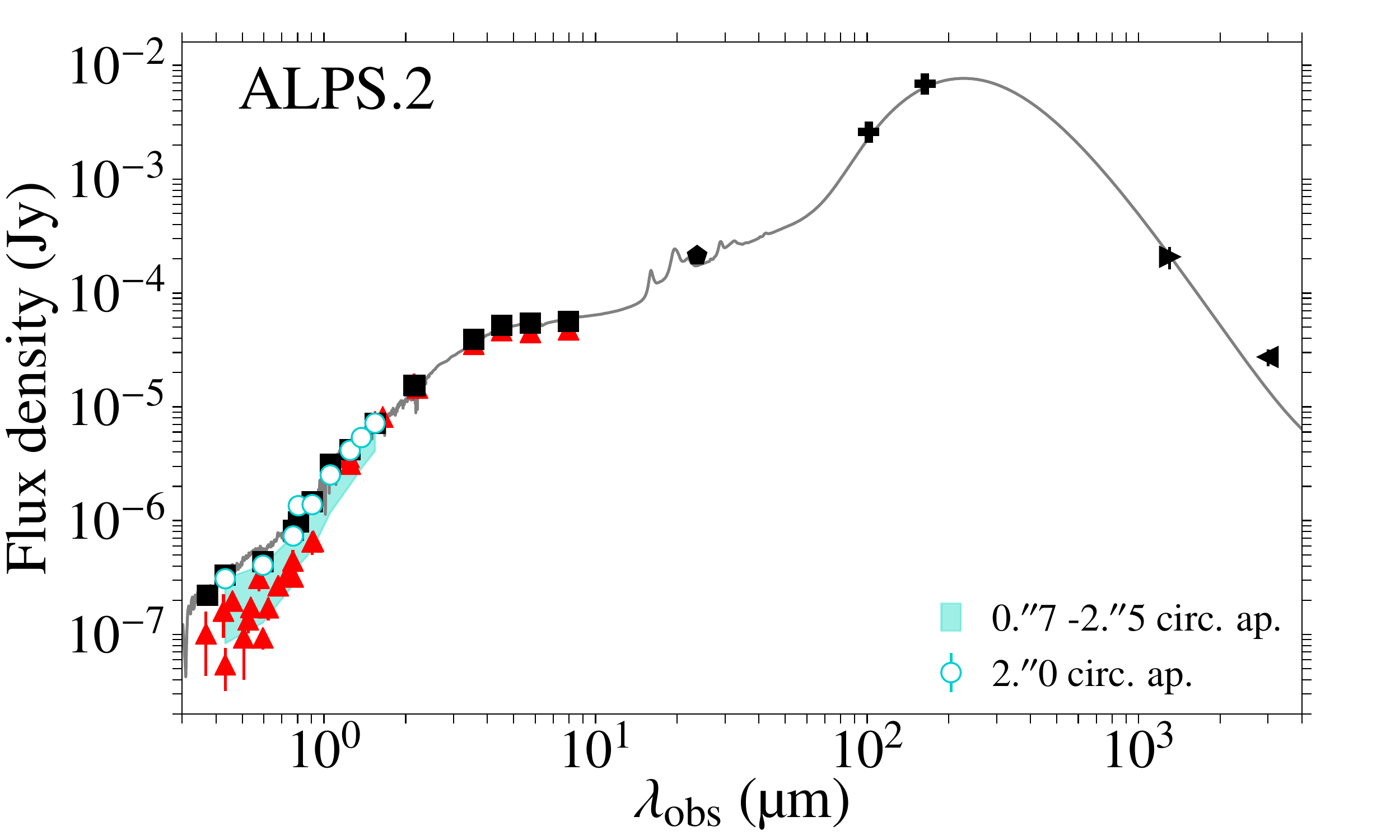}
			\\
			\includegraphics[width=0.48\textwidth, trim={0.1cm 0.cm 0.5cm 0.1cm},clip]{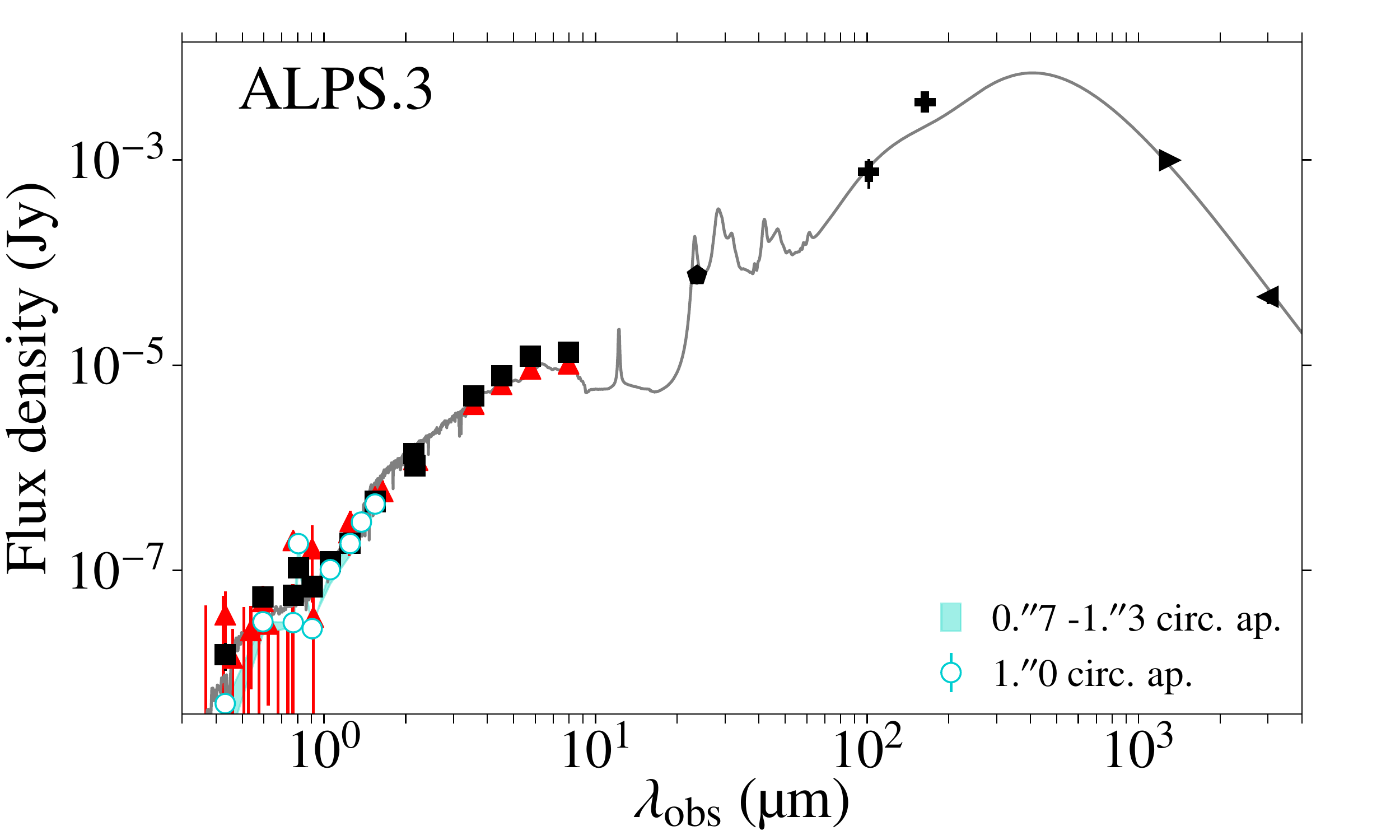}
			\caption{ Photometry used for the SED modeling (black filled symbols) and best-fit solutions (black lines) for our three sources. The UV-IR photometry of \cite{2013ApJS..207...24G} (black squares) and \cite{2014ApJS..214...24S} (red triangles) are also compared (see Appendix \ref{sec:sed_analysis}). Compared to the \cite{2014ApJS..214...24S} photometry, we find the \cite{2013ApJS..207...24G} \ac{hst} photometry to be systematically larger for the measurements from the rest-frame UV-optical (particularly at 1.6\,\textmu m), consistent with what we measure from apertures enclosing the extent of the source emission (cyan circles representing the flux measured from the XDF images within circular apertures encompassing the full extent of source emission). This is especially true for the most extended source, ALPS.2. Also shown are the \emph{Spitzer}, \emph{Herschel} and ALMA continuum data used to constrain the SEDs (as decsribed in the legend). \label{fig:ex_sed}}
		\end{figure}

	\subsection{Stellar Masses and SFRs} 
		\label{sub:stellar_masses}

		We rely on \ac{sed} modeling to infer the stellar masses and SFRs of our sources. As part of the UDF, our sample has been observed with a wide range of ground- and space-based observatories. To trace the extent of the stellar continuum, we rely on the \ac{hst}/WFC3 F160W images from the XDF data release, described in \cite{2013ApJS..209....6I}\footnote{\url{https://archive.stsci.edu/prepds/xdf/}}. Given our small sample size, and use of the \ac{hst}/WFC3 F160W Band to trace the stellar continuum, we revisit the accuracy of the \ac{hst} photometry in Appendix \ref{sec:sed_analysis}. 

		We select the following sets of data for the SED analysis. For the UV-NIR photometry we use the values from the \cite{2013ApJS..207...24G} catalogue, which are extracted from large enough apertures to enclose the full extent of the emission for the extended sources analysed here (see Appendix \ref{sec:sed_analysis}). In addition, we use the deblended \emph{Spitzer}/MIPS 24\,\micron\ photometry from \cite{2014ApJ...795..104W} and the deblended \emph{Herschel}/PACS 100\,\micron\ and 160\,\micron\ data from \cite{2011A&A...533A.119E}. We take the maximum of the local and simulated noise levels as the uncertainties for the \emph{Herschel} data. We also use the 1.3\,mm dust continuum measurements of \cite{2017MNRAS.466..861D} and \cite{2019ApJ...882..139G} and the 3\,mm continuum limits presented in \cite{2019ApJ...882..139G}.  

		We model the photometry using two, adapted, versions of the high-redshift extension to the SED-fitting algorithm {{\sc{MAGPHYS}}} \citep[][]{2008MNRAS.388.1595D,2015ApJ...806..110D}. To account for the impact of obscuration and high SFRs on the attenuation, we apply the adaptation to the high-redshift extension of the {\sc{MAGPHYS}} code, developed by \cite{2019ApJ...882...61B}. We find that this adapted version significantly reduces the fitting residuals, particularly for ALPS.3 (see Appendix \ref{sec:sed_analysis}). Thus, we quote the fitting parameters from this adaptation of {\sc{MAGPHYS}} for ALPS.1 and 3 in Table~\ref{tab:derived_prop}. Despite the more sophisticated dust treatment, the IRAC and PACS photometry for ALPS.2 remain poorly fit by the adapted version of {\sc{MAGPHYS}}. This is likely due to the presence of an AGN, identified from \emph{Chandra} observations \citep[the details of which are described in][]{2017ApJS..228....2L}. Thus, for ALPS.2 we apply the adapted version of {\sc{MAGPHYS}} that accounts for a contribution by AGN to the dust heating \citep{2020ApJ...888...44C}. The inferred stellar mass and SFR are a factor of 2 and 1.6 times smaller, respectively, when the AGN component is considered. We note that this extension does not include the adaptations introduced by \cite{2019ApJ...882...61B}. We provide the parameters from the fit including $\xi_\mathrm{AGN}$, the estimated fraction of the template AGN emission contributing to the total IR luminosity, in Table~\ref{tab:derived_prop}. Based on the variation in the derived stellar masses and SFRs for different sets of photometry, the inclusion/exclusion of an AGN component and the choice of applied dust attentuation curve, we adopt an uncertainty floor on the stellar masses and SFRs of $\pm\sim0.3$ dex consistent with a factor of $\sim 2$ increase/decrease in stellar mass and factor of $\sim 1.5$ increase/decrease in SFR.

		\begin{table} 
\caption{Derived Properties \label{tab:derived_prop}} 
\begin{tabular}{@{}>{\raggedright}m{0.34\columnwidth}>{\centering}m{0.18\columnwidth}>{\centering}m{0.18\columnwidth}>{\centering\arraybackslash}m{0.18\columnwidth}@{}}
\toprule 
Source &ALPS.1 & ALPS.2 & ALPS.3\\ 
\midrule 
\midrule 
SED fitting$^a$ &  & &\\ 
\midrule 
$M_*$ ($10^{11}$ M$_\odot$)  & 1.9 $^{+1.9}_{-1.0}$  & 1.1 $^{+2.2}_{-0.6}$  & 3.0 $^{+3.0}_{-1.5}$ \\ 
$M_\mathrm{dust}^\mathrm{SED}$ ($10^{8}$ M$_\odot$) & 1.1 $^{+0.3}_{-0.2}$ & 1.6 $^{+0.1}_{-0.5}$ & 7.8 $^{+1.2}_{-1.0}$\\ 
SFR (M$_\odot$ yr$^{-1}$)  & 54 $^{+27}_{-13}$ & 48 $^{+24}_{-12}$ & 98 $^{+89}_{-25}$\\ 
sSFR (Gyr$^{-1}$)  & 0.3 $^{+0.1}_{-0.1}$  & 0.4 $^{+0.1}_{-0.8}$  & 0.3 $^{+0.7}_{-0.4}$ \\ 
sSFR/sSFR$_\mathrm{MS}$$^b$ (Schreiber+2015)  & 0.4 $\pm$ 0.4 & 0.4 $\pm$ 0.4 & 0.2 $\pm$ 0.2\\ 
$\xi_\mathrm{AGN}$  & & 0.15 $^{+0.04}_{-0.04}$  & \\ 
\midrule 
\multicolumn{4}{@{}l}{Flux densities, applied ratios and derived masses}\\ 
\midrule 
$J_\mathrm{obs}$ & 2 & 2 & 3\\ 
$S_\mathrm{CO} \Delta v$ (mJy km s$^{-1}$)   & 600 $\pm$ 130 & 560 $\pm$ 80 & 560 $\pm$ 70\\ 
$L^\prime_\mathrm{CO (J_\mathrm{obs} - (J_\mathrm{obs}-1))}$ \\ ($10^{10}$ K km s$^{-1}$ pc$^{2}$)   & 1.6 $\pm$ 0.3 & 1.7 $\pm$ 0.2 & 2.1 $\pm$ 0.3\\ 
$r_{J_\mathrm{obs}, 1}$  & 0.8 $\pm$ 0.2 & 0.8 $\pm$ 0.2 & 0.79 $\pm$ 0.21\\ 
$L^\prime_\mathrm{CO(1-0)}$ \\ ($10^{10}$ K km s$^{-1}$ pc$^{2}$)   & 2.0 $\pm$ 0.3 & 2.2 $\pm$ 0.2 & 2.6 $\pm$ 0.3\\ 
$\alpha_\mathrm{CO}$ $^c$ \\ $(\mathrm{\Msun / (\mathrm{K}\, \kms \, \mathrm{pc}^{2})}$)    & 3.9 $\pm$ 0.4 & 4.2 $\pm$ 0.4 & 4.1 $\pm$ 0.4\\ 
$M_\mathrm{mol}$ ($10^{11}$ M$_\odot$)  & 0.8 $\pm$ 0.3 & 0.9 $\pm$ 0.3 & 1.1 $\pm$ 0.3\\ 
$S_\mathrm{1.3\,mm}$ ($\mathrm{\mu}$Jy)  & 240 $\pm$ 70 & 400 $\pm$ 100 & 860 $\pm$ 120\\ 
$M_\mathrm{dust}^\mathrm{1.3\,mm}$ ($10^{8}$ M$_\odot$)  & 2.4 $\pm$ 0.7 & 4.0 $\pm$ 1.0 & 7.6 $\pm$ 1.1\\ 
\midrule 
\multicolumn{4}{@{}l}{Inferred mass ratios}\\ 
\midrule 
$M_\mathrm{mol}/M_\mathrm{dust}^\mathrm{1.3\,mm}$  & 320 $\pm$ 150 & 230 $\pm$ 90 & 140 $\pm$ 50\\ 
$M_\mathrm{mol}/M_* $  & 0.4 $\pm$ 0.2 & 0.8 $\pm$ 0.5 & 0.4 $\pm$ 0.2\\ 
\bottomrule 
\end{tabular} 
\begin{itemize}[leftmargin=*] 
{\item[$^a$] For properties inferred from the SED fitting we quote the median values with uncertainties representing the 16th and 84th percentiles. For further calculations and for the analysis described in the text, we adopt an uncertainty floor of $\pm 0.3$ dex on the values of $M_*$ and SFR (based on the systematic uncertainties).}  
 {\item[$^b$] Calculated based on the \cite{2015A&A...575A..74S} best-fit main sequence. Uncertainties on the MS offset are calculated based on the systematic uncertainties in the SFRs and stellar masses and the uncertainty of the MS function. }  
 {\item[$^c$] CO(1-0)-to-molecular gas conversion calculated based on the metallities inferred via the stellar mass-metallicity relation parameterised in Equation (12a) of \cite{2015ApJ...800...20G} and applying Equation (2) of \cite{2018ApJ...853..179T}. } \end{itemize} 
\end{table}


		To place our sources in the context of the main sequence, at the measured redshift of each source, we apply the best-fit MS relation of \cite{2015A&A...575A..74S}. We choose this MS for consistency with the HUDF and ASPECS parent samples shown in \cite{2020arXiv200604284A} (see their Figure 6), which we also compare to in Figure \ref{fig:selection}.  We note that the exact shape of the MS, particularly for the large stellar masses of our three galaxies, is still under debate \citep[the functional forms of e.g.][differ]{2012ApJ...754L..29W,2014ApJ...795..104W,2014ApJS..214...15S,2015A&A...575A..74S,2018A&A...619A..27B,2020arXiv200613937L}. Our sources range from having sSFRs consistent with the MS to being slightly below the MS, with specific star formation rates relative to the main sequence of sSFR/sSFR$_\mathrm{MS} = 0.2-0.4$, based on the \cite{2015A&A...575A..74S} MS. This places our three galaxies in a regime that has been scarcely sampled to date (e.g. \citealt{2015ApJ...809..175B} have resolved CO observations for two, MS sources at $z\sim 2$). Moreover, based on the derived flux densities, our sources appear to have large gas mass fractions $\mathrm{M_{mol}}/\Mstar = 0.4-0.8$ and gas-to-dust ratios typical of MS galaxies and SMGs, i.e. $\mathrm{M_{mol}}/\mathrm{M_{dust}} \sim 100-300 $ (for this comparison we use the dust mass inferred from the 1.3 mm continuum).  

	

	\begin{figure*}
		\centering
		\includegraphics[width=\textwidth, trim={1.2cm 0.8cm 1.2cm 0.8cm},clip]{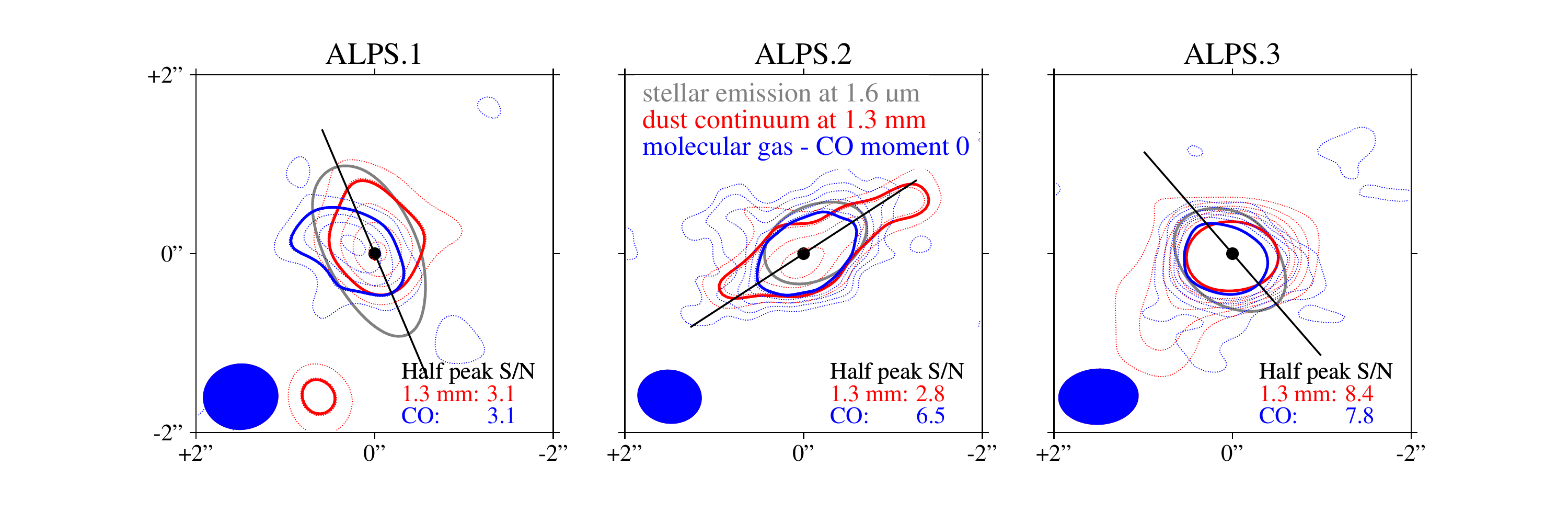}
		\caption{Comparison of the half-peak emission contours (solid lines) for the stellar emission at 1.6\, \textmu m (grey), the 1.3 mm dust-continuum emission (red) and the CO line emission (blue), which probes the molecular gas. Both the 1.3\,mm dust continuum and \ac{hst} F160W maps have been convolved to the resolution of the CO emission. Each panel shows a different source, labeled at the top. The dotted contours depict the CO (blue) and convolved 1.3\,mm continuum (red) data, starting at $+2\sigma$ and increasing in steps of $+1\sigma$. The ratio of the half-peak value to the rms (half peak S/N) for both the 1.3\,mm and CO data are listed on the bottom right of each panel.   The black filled circle and line represent the kinematic centre and position angle used to extract the major axis profiles in the next figure. The \ac{hst} astrometry has been corrected according to the offsets measured by \cite{2020arXiv200503040F}. \label{fig:overlays}}
	\end{figure*}

	\begin{figure}[h]
		\centering
		\includegraphics[width=0.44\textwidth, trim={0cm 0.5cm 0cm 0.2cm},clip]{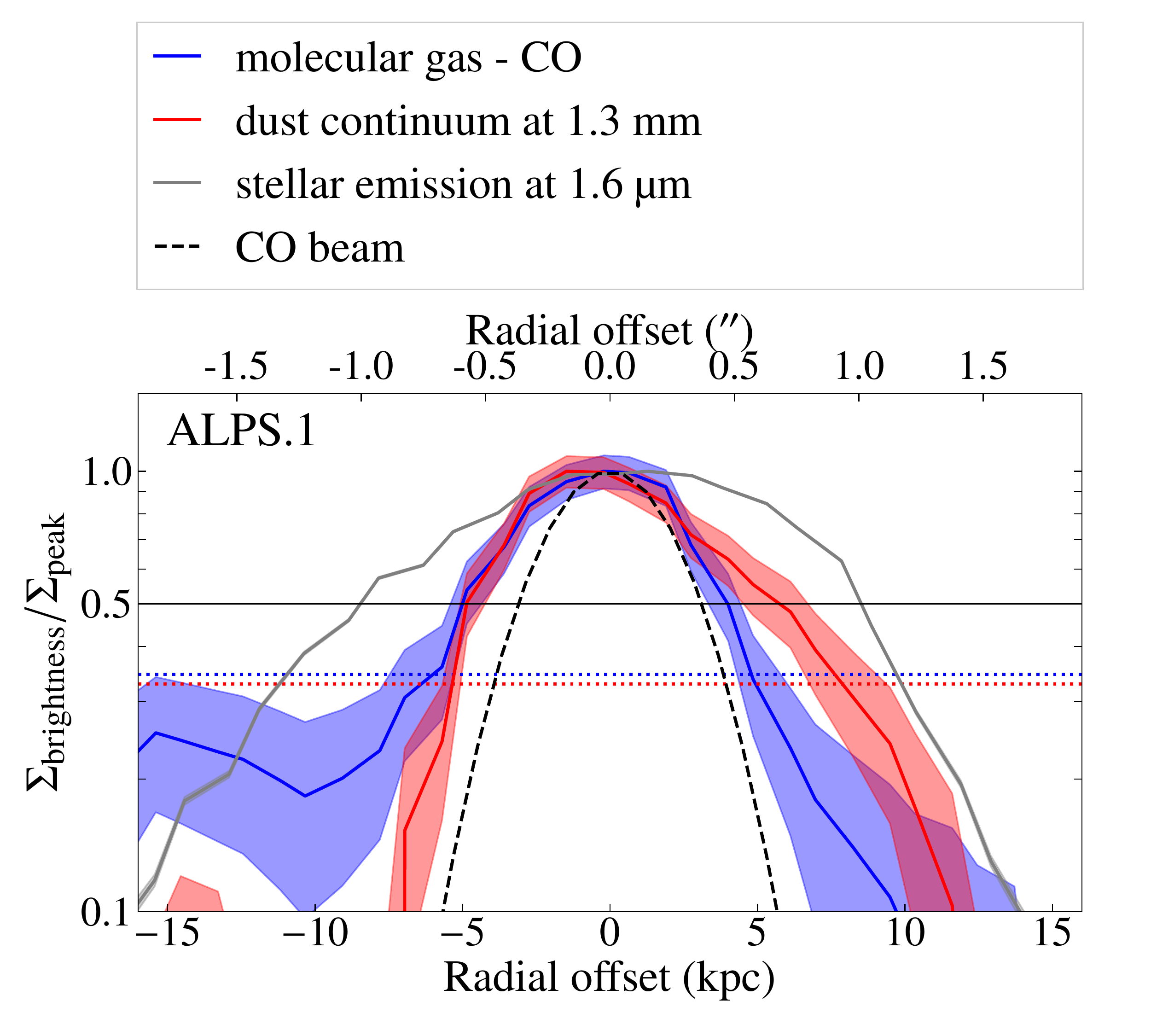}
		\\
		\includegraphics[width=0.44\textwidth, trim={0cm 0.5cm 0cm 6.cm},clip]{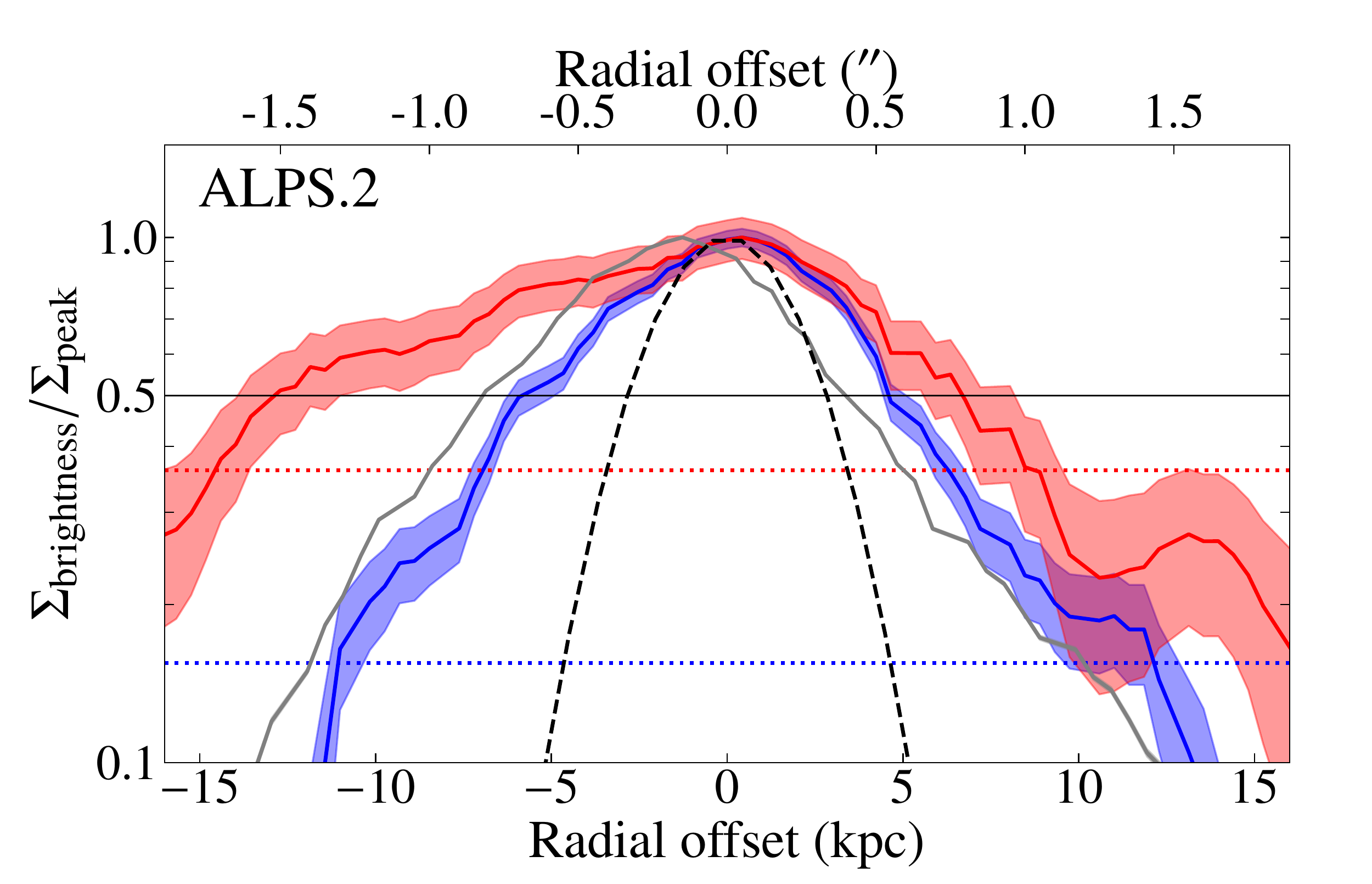}
		\\
		\includegraphics[width=0.44\textwidth, trim={0cm 0.5cm 0cm 6.cm},clip]{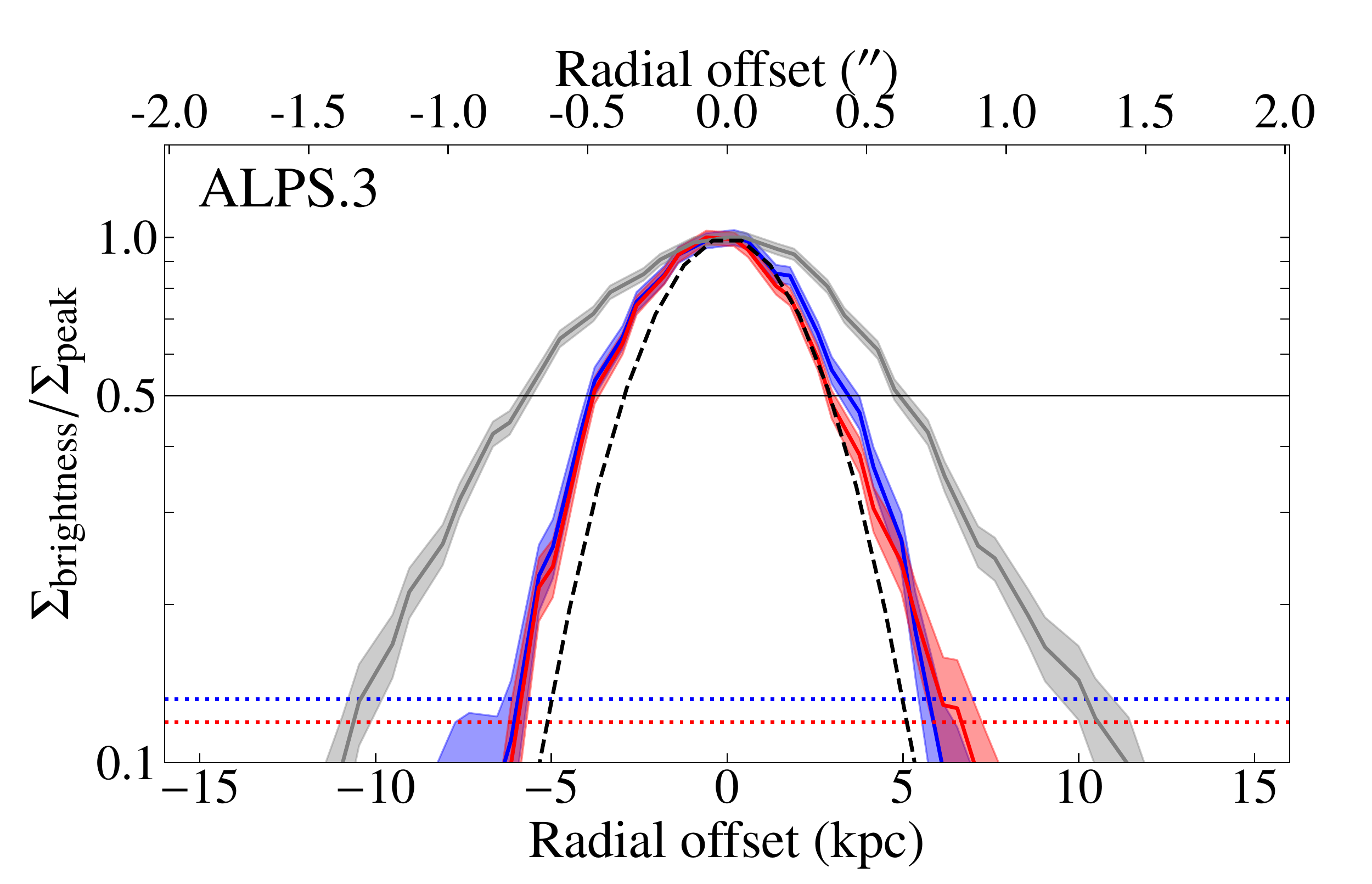}
		\caption{Comparison of the normalized surface brightness profiles ($\Sigma_\mathrm{brightness}/\Sigma_\mathrm{peak}$) of stellar emission from the F160W image (grey), 1.3 mm dust continuum (red) and CO emission, from the moment 0 map (blue) along the major axes of each source (labeled at top left of each panel). The horizontal black line of each panel indicates half the maximum surface brightness, thereby allowing the FWHM of the three profiles to be compared. The $2\sigma$ level relative to the peak surface brightness of the CO and dust-continuum emission are indicated by the blue and red dotted lines, respectively. The major axes, from which these profiles are extracted, are shown in Figure~\ref{fig:overlays}. The \ac{hst} astrometry has been corrected according to the astrometric offsets provided by \cite{2020arXiv200503040F}.  \label{fig:major_axis_profiles}}
	\end{figure}



\begin{table} 
\caption{H$_2$ Mass and Column Sensitivities \label{tab:sensitivities}} 
\begin{tabular}{@{}>{\raggedright}m{0.34\columnwidth}>{\centering}m{0.18\columnwidth}>{\centering}m{0.18\columnwidth}>{\centering\arraybackslash}m{0.18\columnwidth}@{}}
\toprule 
Source & ALPS.1 & ALPS.2 & ALPS.3\\ 
\midrule 
\multicolumn{4}{@{}l}{rms values of the ...}\\ 
\midrule 
1.3\,mm map (\textmu Jy)  & 15 & 18 & 14\\ 
CO moment-0 map (mJy km s$^{-1}$) & 49 & 20 & 25\\ 
\midrule 
\multicolumn{4}{@{}l}{H$_2$ mass limit per beam (10$^8$ M$_\odot$ beam$^{-1}$)}\\ 
\midrule 
Dust-based$^a$  & 28\,$\pm$\,14 & 34\,$\pm$\,17 & 24\,$\pm$\,12\\ 
CO-based$^b$  & 55\,$\pm$\,\phantom{0}6 & 28\,$\pm$\,\phantom{0}3 & 48\,$\pm$\,\phantom{0}5\\ 
\midrule 
\multicolumn{4}{@{}l}{H$_2$ column density limit per beam ($10^{21}$ cm$^{-2}$ beam$^{-1}$)$^{c}$}\\ 
\midrule 
Dust-based  & 2.7\,$\pm$\,1.4 & 4.6\,$\pm$\,2.3 & 2.9\,$\pm$\,1.4\\ 
CO-based  & 5.3\,$\pm$\,0.5 & 3.9\,$\pm$\,0.4 & 5.7\,$\pm$\,0.6\\ 
\bottomrule 
\end{tabular} 
\begin{itemize}[leftmargin=*] 
{\item[$^{a}$] Calculated assuming a gas-to-dust mass ratio of $200\,\pm\,100$, with the dust emission described by a MBB with $T=25$\,K, $\beta=1.8$.}  
 {\item[$^b$] Calculated based on the $\aco$ in Table \ref{tab:derived_prop}.}  
 {\item[$^c$] Calculated using the CO beam size in Table \ref{tab:source_prop}.}  
 {\item[$^*$] The sensivity estimates are based on $1\sigma$ (i.e. applying the rms values at the top of the table). } \end{itemize} 
\end{table}

\section{Galaxy Size Analysis} 
	\label{sec:source_size_analysis}

		\subsection{Sensitivity Comparison} 
			\label{sub:sensitivities}

			Before comparing the spatial distribution of the dust-continuum and CO emission, we investigate how sensitive our ALMA data is to to the molecular gas. We estimate the minimum H$_2$ mass and column densities that are observable based on our 1.3\,mm continuum and CO moment-0 maps, comparing values at a matched resolution given by the CO moment-0 map beam size. To estimate the CO-based H$_2$ mass limit per beam, we take the rms of the moment-0 maps and scale to a line luminosity limit using the $\aco$ conversion factors in Table \ref{tab:derived_prop}. We account for the solid angle of the beam to derive the column density limit. For the dust-based values, we take the rms of the dust-continuum maps and scale these values to the beam size of the CO moment-0 maps. We convert the rms per beam to an inferred H$_2$ mass per beam by assuming a modified blackbody with the same assumptions on the temperature, emissivity index and opacity described in Section \ref{sub:flux_densities}. We apply a gas-to-dust ratio of $200\pm100$, consistent with what we measure for our sample (Table \ref{tab:derived_prop}). Based on these assumptions we derive the sensitivities in Table \ref{tab:sensitivities}. Note that here we compare the values for H$_2$ only, whereas we consider the total molecular gas masses (which include He) in the rest of the text.

			The sensitivities calculated here, multiplied by a suitable S/N, represent the minimum observable column density assuming that the total area of the beam is covered by source emission (i.e. a beam filling factor of unity). In the case of a spiral arm, the actual CO column would only fill a fraction of the beam, e.g. if a spiral arm only covers a tenth of the beam area, then the actual column density of molecular gas that our data are sensitive to is ten times what is quoted here. To detect the molecular gas at $3\sigma$, the column densities would have to be $\sim 10^{22} \mathrm{cm}^{-2}$ (modulo the beam filling factor). For local disks, the typical column densities of the bar and spiral arms are $10^{21} - 10^{22} \mathrm{cm}^{-2}$ whereas the column densities measured for the central starbursts are up to an order of magnitude higher, e.g. the column density of the nuclear region of NGC 253 is $2 \times 10^{23} \times \mathrm{cm}^{-2}$ \citep{2008A&A...490...77W}. This difference in column densities leads to a significant contrast between the nuclear region and brightest parts of the disks (and bar), i.e. factor of $20-50\times$ in the observed flux densities. 

			Based on the estimated sensitivities, our CO and dust-continuum observations appear to be sensitive to approximately the same column of molecular gas. The values estimated for the CO and continuum data are consistent within the uncertainties (based on the above assumptions) for ALPS.1 and 2, whereas for ALPS.3 the dust continuum may be marginally (up to 20\%) more sensitive to the H$_2$ column. Comparing the CO data across our sample, the sensitivity is highest for ALPS.2.  However, the continuum data of our sample have similar sensitivities, with slightly lower inferred H$_2$ mass and column density limits per beam for ALPS.1 and 3 versus ALPS.2. The comparison between CO and dust-based H$_2$ column sensitivities is based on the assumptions that the dust temperature and gas-to-dust ratio remain constant across the disk. Variations in the temperature or gas-to-dust ratio would serve to decrease the sensitivities of the dust emission, compared to CO, at larger radii.


		\subsection{Qualitative Comparison of Spatial Extents} 
			\label{sub:qualitative_comaprison}

			Our aim is to quantify the sizes of the rest-frame optical, CO and dust-continuum emission. However, to aid the interpretation of our data, we begin with a qualitative comparison.  In Figure~\ref{fig:overlays}, we compare the observed galaxy extents at half the peak surface brightness (solid lines) of the rest-frame optical, dust continuum and CO emission (with both the F160W and Band 6 dust continuum convolved to the resolution of the CO moment-0 maps). The F160W maps have been corrected for the offset of the \ac{hst} astrometry measured by \cite{2020arXiv200503040F} (for which the median offset is $-96$\,mas in right ascension and $+261$\,mas in declination and the additional local offsets for our sources differ from the medians by $\leq 40$\,mas). We list the S/N at half the peak surface brightness for the CO and 1.3\,mm continuum in the bottom right corner of each panel. For ALPS.1, the CO and 1.3\,mm continuum half-peak values are barely greater than $3\sigma$. For ALPS.2, the half-peak value of the 1.3\,mm emission is below $3\sigma$. In contrast, for the 1.6\,\textmu m data, the S/N at half the peak value is at $>50$ for each source.

			We compare the normalized surface brightness profiles extracted along the entire major axis (see Figure ~\ref{fig:major_axis_profiles}). The normalized, major axis cut illustrates how the steepness of the surface brightness profiles  compare and indicates a potential difference in the peak positions (for the three types of emission). We show the beam-convolved, major axis profiles because the accurate construction and interpretation of the radial profiles for the CO and 1.3\,mm continuum emission was hindered by the poor S/N and resolution. To indicate how well our data are resolved, we compare the CO beam (dashed black lines) to the major axis profiles and show the $2\sigma$ level normalized to the peak flux of the CO and 1.3\,mm continuum emission (dotted blue and red lines).

			From Figures \ref{fig:overlays} and \ref{fig:major_axis_profiles} it appears that for ALPS.1, little CO or dust-continuum emission is recovered, at $>2\sigma$, on scales greater than the beam size. For ALPS.2, a significant fraction of CO and dust-continuum emission appears to be on scales larger than the beam size. In contrast, for ALPS.3, the dust and CO emission are centrally concentrated, with scarcely any emission apparent on scales greater than the image resolution. We quantify the fraction of flux within the beam area (i.e. consistent with a point source) in Section \ref{sub:measuring_galaxy_sizes}. The stellar-continuum emission from ALPS.1 and 3 appears to stem from a significantly larger region than the CO or dust continuum. However, for ALPS.2, the stellar emission follows a similar profile to the CO emission, with the dust continuum exhibiting a flatter, extended profile, albeit at a S/N$<3$.


		\begin{table} 
\caption{Inferred Source Sizes \label{tab:source_sizes}} 
\begin{tabular}{@{}>{\raggedright}m{0.34\columnwidth}>{\centering}m{0.18\columnwidth}>{\centering}m{0.18\columnwidth}>{\centering\arraybackslash}m{0.18\columnwidth}@{}}
\toprule 
Source &ALPS.1 & ALPS.2 & ALPS.3\\ 
\midrule 
\midrule 
\multicolumn{4}{@{}l}{Structural parameters from \cite{2012ApJS..203...24V}$^a$}\\ 
\midrule 
r$_{1/2}^{\mathrm{F160W}}$ (kpc) & 7.46 $\pm$ 0.02 & 8.28 $\pm$ 0.03 & 4.84 $\pm$ 0.15\\ 
S\'{e}rsic index & 0.48 $\pm$ 0.00 & 3.04 $\pm$ 0.02 & 0.86 $\pm$ 0.06\\ 
Position angle (degrees)  & 23.2 $\pm$ 0.1 & 302.5 $\pm$ 0.1 & 220.9 $\pm$ 1.1\\ 
Axis ratio  & 0.247 $\pm$ 0.002 & 0.452 $\pm$ 0.002 & 0.458 $\pm$ 0.013\\ 
\midrule 
\midrule 
\multicolumn{4}{@{}l}{Half-light radii measured here using \texttt{GALFIT} $^b$} \\ 
\midrule 
r$_{1/2}^{\mathrm{F160W}}$ (kpc) & 7.9 $\pm$ 0.8 & 8.0 $\pm$ 0.8 & 4.9 $\pm$ 0.5\\ 
S\'{e}rsic index (F160W) & 0.6 $\pm$ 0.1 & 2.1 $\pm$ 0.2 & 0.7 $\pm$ 0.1\\ 
$r_{1/2}^{\mathrm{CO}}$ (kpc) & 5.8 $\pm$ 1.7 & 5.5 $\pm$ 0.8 & $<$ 3.4\\ 
$r_{1/2}^{\mathrm{dust}}$ (kpc) & 3.9 $\pm$ 1.2 & 9.4 $\pm$ 1.4 & $<$ 1.2\\ 
\midrule 
\bottomrule 
\end{tabular} 
\begin{itemize}[leftmargin=*] 
{\item[$^a$] Used as priors for the \texttt{GALFIT} fitting.}  
 {\item[$^b$] Measurements based on an exponential surface brightness profile. } \end{itemize} 
\end{table}

		\subsection{Measuring Galaxy Sizes} 
			\label{sub:measuring_galaxy_sizes}

			To quantify the source sizes, we attempt various fitting methods.  For the interferometric data (i.e. the CO and dust continuum data) there are two possible approaches;  fitting the maps (i.e. the image plane), or, directly fitting the visibilities (i.e. the $uv$-plane). For both approaches we use the axis ratios and position angles determined by \cite{2012ApJS..203...24V} from the F160W images and provided in Table~\ref{tab:source_sizes}, as initial estimates.  

			\subsubsection{Image-plane Analysis}

\begin{figure*}
					\centering
					\includegraphics[width=\textwidth, trim={0.5cm 0.5cm 1.2cm 0.5cm},clip]{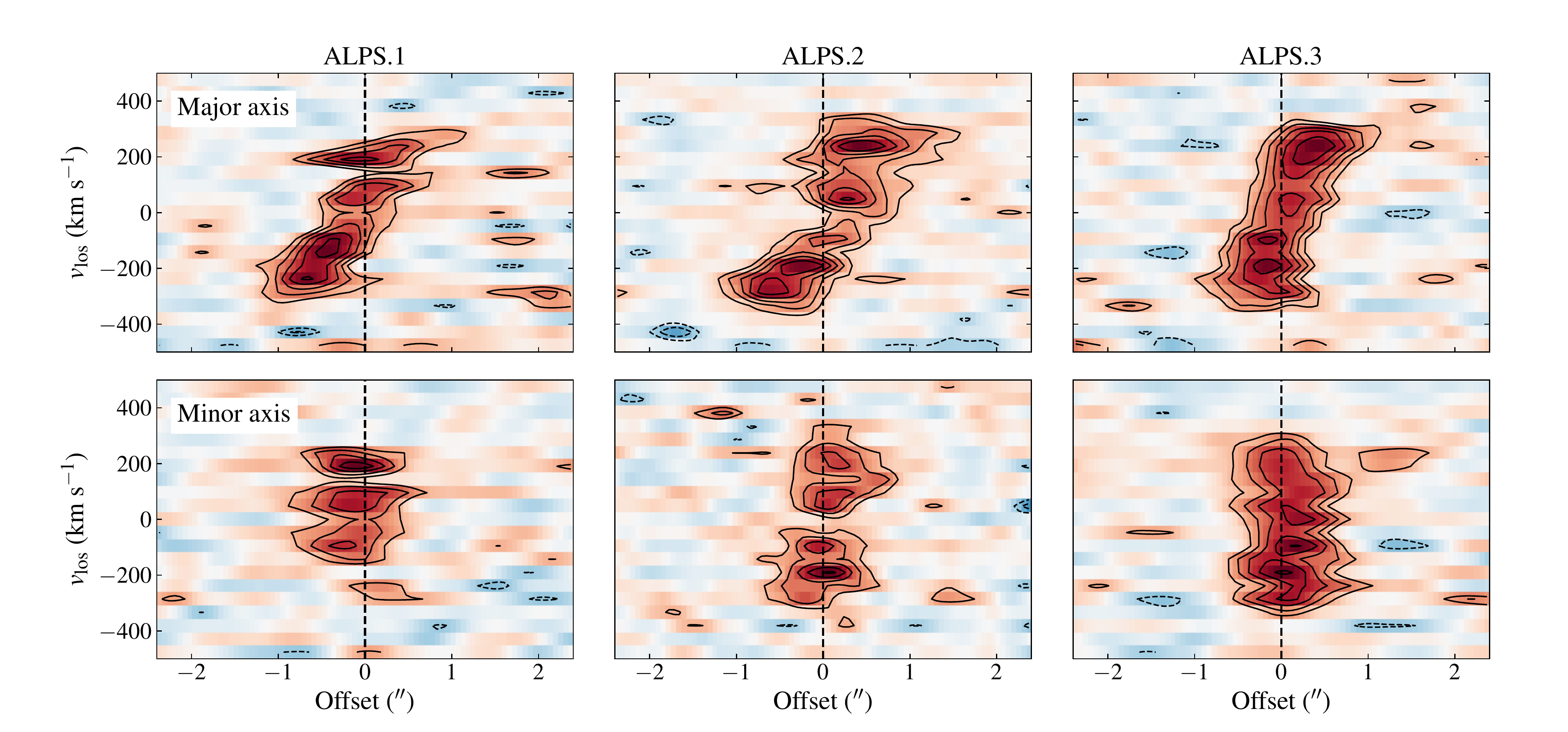}
					\caption{ Position-velocity (pV) diagrams extracted over $\pm\, 500\, \kms$ within $0\as3$ wide slits along the rotation axis (where the position angle is taken from the kinematic modeling of the CO data, described in Section \ref{sub:kinematic_modeling}) using \texttt{CASA}'s \texttt{impv}. Top row: pV diagram along along major axis. We see clear evidence of rotation in all three galaxies. Bottom row: pV diagram along minor axis. The flux density is represented by the linear, blue-to-red colorscale where white represents the zero level. Contours start at $\pm 2\sigma$ and proceed in steps of $\pm 1\sigma$.   \label{fig:pv_diagrams}}
				\end{figure*}

				Before performing the image-plane analysis, we check whether our ALMA data have sufficient sensitivity and spatial resolution to estimate source sizes. To this end, we fit the emission of each source with a model point source. We check the residuals of the image-plane fit (subtracting the beam-convolved point source model from the image) and compare the point source fluxes to the measured flux densities. For ALPS.1 and 2 we recover remaining, residual structure (at $>2\sigma$) and find that the flux densities of the point source fits are $ 40 \pm 20\%$ lower than our measured values. Conversely, for ALPS.3, we recover no significant residual CO or continuum emission at a spatial scale greater than that of the sythesized beams and find that the flux densities of the point source models are consistent with the measured values. Thus, for ALPS.3, we place $2\sigma$ upper limits on the half-light radii that can be measured from the moment-0 CO and dust continuum maps. Because the emission of ALPS.3 is indistinguishable from the beam, we quantify the $2\sigma$ upper limit based on the convolution of two Gaussian profiles, via, 
				\begin{align}
					FWHM_\mathrm{true}^2 < & (FWHM_\mathrm{beam} + 2 \delta FWHM_{obs})^2 \nonumber \\
					& - FWHM_\mathrm{beam}^2 ,
				\end{align}
				where $FWHM_\mathrm{true}$ is the intrinsic FWHM of the galaxy, assuming a Gaussian profile, and $\delta FWHM_\mathrm{obs}$ is the error on measuring the FWHM based on the rms of the map. The values are given in Table~\ref{tab:source_sizes}. Although ALPS.3 is not sufficiently resolved in the CO moment-0 map, the 3D data indicates that ALPS.3 is marginally resolved, with the location of the emission shifting for different velocity channels (as indicated by the position-velocity diagram in Figure \ref{fig:pv_diagrams}).
				
				Having established that the CO line and 1.3\,mm continuum emission of ALPS.1 and 2 are sufficiently resolved, we attempt to quantify the potential asymmetries visible in Figures \ref{fig:overlays} and \ref{fig:major_axis_profiles}. Following the second equation of Section 3.2 in \cite{2000ApJ...529..886C}, we calculate the asymmetry parameter (see also \citealt{2003ApJS..147....1C}). Note that a value of 0 corresponds to a galaxy that is perfectly symmetric whereas a value of 1 indicates that a galaxy is completely asymmetric. For ALPS.1, we derive asymmetry parameters of $A=0.5\pm0.2$ and $A = 0.7\pm0.1$ for the CO and 1.3\,mm continuum emission, respectively. Similarly, for ALPS.2, we derive asymmetry parameters of $A=0.6\pm0.1$ and $A = 0.7\pm0.1$ for the CO and 1.3\,mm continuum emission. Although this indicates that the observed emission is somewhat asymmetric, for both types of emission, we note that the exact noise structure and resolution heavily bias these values in ways that are challenging to quantify. Nonetheless, the values calculated here indicate that the surface brightness profiles are not well described by symmetric two-dimensional (2D) profiles.

				Despite these potential asymmetries, we estimate the stellar, dust and CO half-light radii by fitting the respective maps with 2D surface brightness profiles via {\sc{GALFIT}} \citep{2002AJ....124..266P,2010AJ....139.2097P}. For each source of emission (stellar, dust and CO), we fit a $4''$ region centred on the source. We consider uniform error maps, with a constant background noise given by the \ac{rms} of the full image. For the input PSFs, we supply images of 2D Gaussian profiles according to the \ac{hst} PSF size and major and minor FWHM of the synthesized beam (of the ALMA data). We allow the position and inclination angles of the best-fit models to be within $\pm 10^\circ$ of the best-fit values from \cite{2012ApJS..203...24V} and constrain the source centres to vary by $\pm\, 0\as2$ of the centres estimated by \cite{2012ApJS..203...24V}.

				We perform multiple tests to check to what extent we can recover the source sizes and S\'{e}rsic profile shapes for different intrinsic sizes, profile shapes and peak signal-to-noise values. Based on sets of simulated data at the resolution and S/N of the observations, we find that the we can accurately constrain the S\'{e}rsic index, $n$, of the 1.6\,\textmu m surface brightness profiles. For all three galaxies, exponential profiles ($n=1$) provide a poor fit to the 1.6\,\textmu m emission \citep[see also][]{2018ApJ...863...56C}.  In contrast to the 1.6\,\textmu m emission, the S\'{e}rsic indices cannot be accurately constrained for the dust and CO emission (based on the S/N and resolution). We therefore keep the S\'{e}rsic index as a free parameter when fitting the F160W emission but fix the S\'{e}rsic index to match an exponential profile ($n=1$) for the CO and dust emission. This choice of exponential profile is motivated by the exponential dust-continuum profiles observed for local \citep[e.g.][]{1998A&A...338L..33H,2007A&A...471..765B,2009ApJ...701.1965M,2011A&A...531L..11B} and high-redshift galaxies \citep{2016ApJ...833..103H,2016ApJ...827L..32B,2017ApJ...834..135T,2018ApJ...863...56C,2019MNRAS.490.4956G}. Fixing the profile shape for the CO and dust continuum has negligible impact on our conclusions as the half-light radii inferred when forcing $n=1$ are consistent, within errors, with the values inferred when $n$ is a free parameter. The same is not true if we fix the 1.6\,\textmu m emission to be exponential. In that case, the half-light radii measured for ALPS.1 and 2 are 20\% larger and 30\% smaller respectively, thanks to the presence of multiple, unobscured, stellar components, visible in the 1.6 \textmu m emission, as well as possible contributions from dust lanes. 

				The uncertainties returned by {\sc{GALFIT}}, based on the uniform error maps, underestimate the uncertainty on the fit parameters. We therefore quote uncertainties based on the distribution of best-fit values for fits to simulated maps. For the F160W emission, the error on the measured half-light radii is $\leq 10\%$ whereas for the CO and dust-continuum emission the uncertainties (not accounting for the uncertainty of the profile shape) are $\sim 20 - 30\%$. 

				The F160W half-light radii measured here are consistent with those of \cite{2012ApJS..203...24V}. The small differences in the best-fit values appear to be due to the deeper XDF data of \cite{2013ApJS..209....6I} used here. The steepness of the stellar surface brightness profiles differs significantly for the three sources. ALPS.1 exhibits the steepest surface brightness profile, with a S\'{e}rsic index of $n\sim 0.5$, equivalent to a Gaussian profile. The unobscured stellar emission of ALPS.3 is best fit by $n\sim 0.8$ (close to the exponential profile, $n=1$), whereas ALPS.2 exhibits the shallowest, unobscured stellar profile with $n>2$. We note that of the sources at $1<z<3$ in the GOODS-South field with measured structural properties a third have profiles with $n>2$ whereas 12\% are best fit by S\'{e}rsic indices $n\lesssim 0.5$. 

				\begin{figure*}
					\centering
					\includegraphics[width=\textwidth, trim={0.5cm 0.cm 0cm 0.cm},clip]{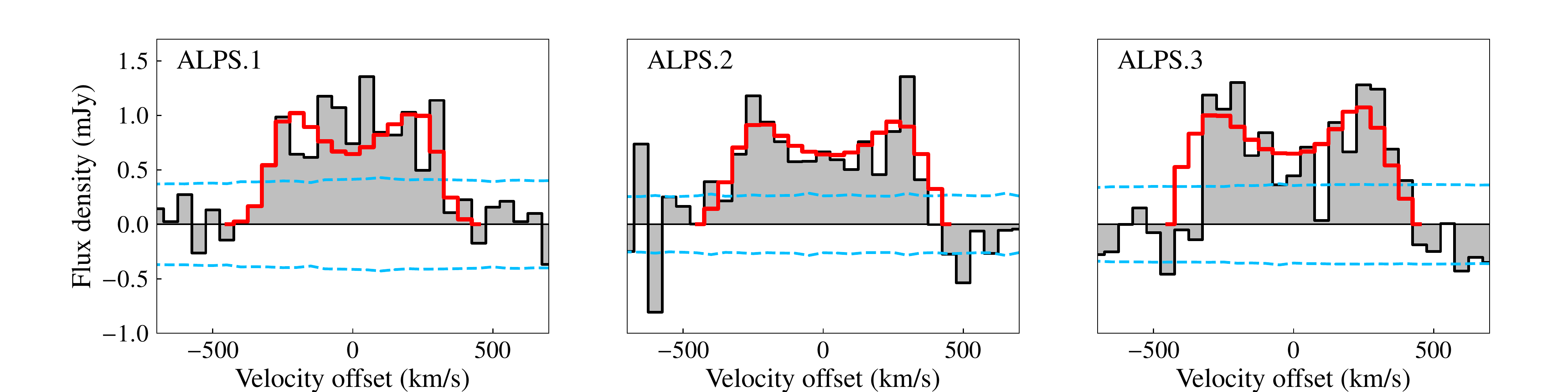}
					\caption{CO spectra of our sources, taken from within 1.5, 2 and 1.5 arcsecond apertures respectively for ALPS.1, 2 and 3. We compare the line profiles from the 3D convolved model of the CO emission, fit using \texttt{qubefit} (in red). The $\pm \sigma$ levels are indicated by the light blue, dahsed lines. All line profiles are inconsistent with a single Gaussian profile. The profiles for ALPS.2 and 3 are clearly double-horned, indicating that the CO emission stems from rotation-dominated disks.  \label{fig:line_profile}}
				\end{figure*}

			\subsubsection{$Uv$-plane Analysis}

				To assess the reliability of our image-plane analysis, we also perform $uv$-plane source fitting using: 1) the {\sc{CASA}}-based, {\sc{uvmultifit}} algorithm \citep{2014A&A...563A.136M} and, 2) the {\sc{GILDAS}}-based\footnote{\url{http://www.iram.fr/IRAMFR/GILDAS}} {\sc{uv\_fit}} algorithm \citep{2005sf2a.conf..721P}. Both algorithms fit the specified Fourier-transformed 2D surface brightness profiles, i.e. the model visibilities, to the measured visibilities by minimizing the $\chi^2$ statistic. We correct for the fact that our sources are not at the centre of the pointings by applying the necessary phase shifts, baseline reprojections and primary beam corrections. These corrections are applied as part of the {\sc{uvmultifit}} algorithm whereas for the second approach, we corrected for the primary beam attenuation and weighting using {\sc{GILDAS}} algorithms before merging the visibilities from the various mosaic pointings for each source. We fit both 2D elliptic Gaussian and exponential profiles but find no significant differences ($<5\%$ in the flux and $<25\%$ in size) between the two profiles. We refer to the exponential fits from hereon for consistency with the image-plane analysis. Additional uncertainties are introduced by the relative weighting of the low- versus high-resolution data sets, with differences of up to 40\% in the inferred sizes (i.e. comparing the fits using both data sets versus the high-resolution set only).

				For all three sources, we find good agreement between the fluxes fit in the $uv$-plane and those measured from the image plane (as described in Section \ref{sub:flux_densities}). Without fixing any parameters, we find that the $uv$-based CO- and 1.3\,mm size estimates for ALPS.1 and 3 are consistent, within $1\sigma$ errors, with the measurements and upper limits from the image plane, whereas for ALPS.2 (owing to the low S/N and the apparently complex morphology) the sizes fit in the $uv$-plane are $50 \pm 30$\% smaller than those fit in the image-plane. However, when we fix the source centre and position angle to the results of our CO kinematic fitting (described in Section \ref{sub:kinematic_modeling}) during the $uv$ fitting, we obtain half-light radii that are consistent with what we measure from the image plane. Although 20\% smaller on average, we conclude that our $uv$- and image-plane analyses are consistent within errors for all three sources, when the source centres and position angles are fixed. Henceforth, we quote the image-plane sizes.

	


\section{Dynamical Analysis} 
	\label{sec:dynamic_analysis}

	\subsection{CO line kinematics} 
		\label{sub:line_kinematics}

		To understand the dynamical properties of our sources, we first analyse their position-velocity (pV) diagrams, shown in Figure~\ref{fig:pv_diagrams}. We create the pV maps from the 50 \kms, CO cubes using {\sc{CASA}}'s {\sc{impv}} task, selecting a $0\as3$ slit along the major axis, which we define using the centres and position angles fit to the CO data (see Section \ref{sub:kinematic_modeling}). We find clear velocity gradients for all three sources (also evident from the moment-1 maps in Figure \ref{fig:qubefit_maps_alps}), suggesting that the bulk of the emission stems from rotation-dominated gas disks. The maximum line-of-sight velocities are $\sim 300\, \kms$, indicating rotation velocities of $310-340\, \kms$ (based on the inclinations inferred from the \ac{hst} data). 

		We also assess the integrated CO spectra of our sample. The spectra from the ASPECS LP data, are shown in Figure 2 of \cite{2019ApJ...882..136A} and Figure 8 of \cite{2019ApJ...882..139G}. We also provide the spectra of our three sources, based on the combined ASPECS LP and ALPS data, in Figure \ref{fig:line_profile}, comparing the line profile predicted from the dynamical modeling (described in Section \ref{sub:kinematic_modeling}). The observed shape of the line profile is governed by multiple effects including the inclination, velocity dispersion, total mass (related to the stellar and gas mass distributions) and steepness of the gas surface density profile \citep{2014AJ....147...96D}. Distinct, double-horned profiles originate from gas that is supported by rotation in a disk, and for which the rotation velocity increases sharply at small radii and flattens at larger radii. ALPS.2 and 3 both display double-horned profiles, indicating that the emission from each source stems from a rotation-dominated disk.  Such profiles are commonly observed for local disk galaxies using HI, which traces the gas disk to large radii \citep[e.g.][]{2010MNRAS.403..683C}.

		About half of the SMGs observed at sufficient S/N exhibit double-peaked lines profiles \citep[e.g.][]{Birkin2020}. Double-horned profiles are more commonly observed for massive star-forming galaxies with extended molecular gas disks (such as our sample) \citep[e.g.][]{2013ApJ...768...74T}. The flatter-than-Gaussian profiles observed for some high-redshift galaxies are sometimes interpreted as further evidence that the gas is more turbulent than in local galaxies. However, for many observed sources it is also possible that the observed emission stems from a small, central region that doesn't probe the maximum, rotation velocity (i.e. where the velocity curve flattens). Although we infer the velocity dispersions and maximum rotation velocities in the next section, we note here that the flat CO line profile of ALPS.1, compared to the separate peaks observed for ALPS.2 and 3, may be the result of a steeper surface density profile (i.e. as indicated by the smaller S\'{e}rsic index of the fit to the stellar emission) or the fact that the CO observations, for this source, are less sensitive to the flat part of the rotation curve. 

		\begin{figure*}[!htb]
			\centering
			\captionsetup{list=off}
			\includegraphics[width=0.84\textwidth, trim={0.cm 0.cm 0.cm 0.cm},clip]{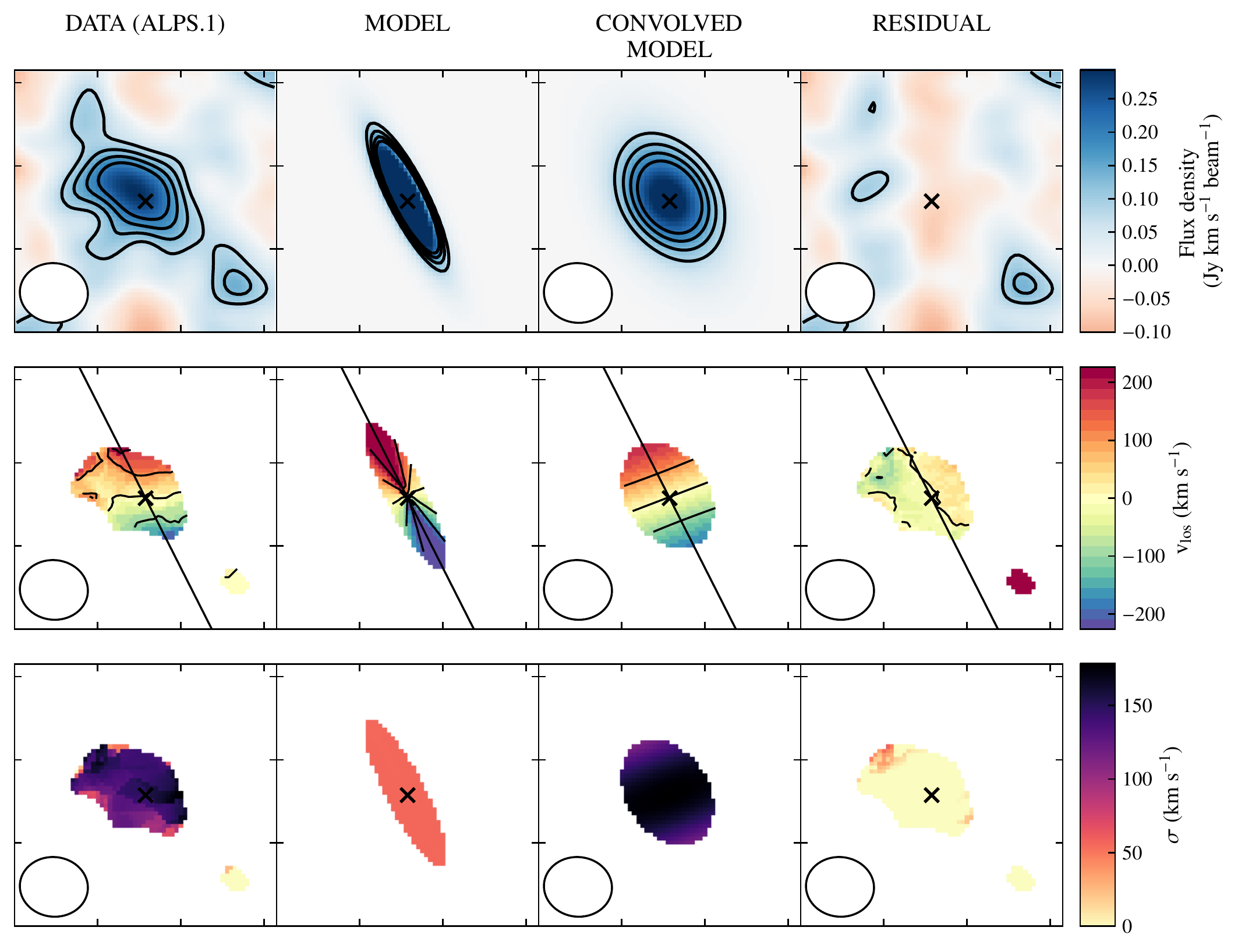}
			\\ 
			\includegraphics[width=0.84\textwidth, trim={0.cm 0.cm 0.cm 0.cm},clip]{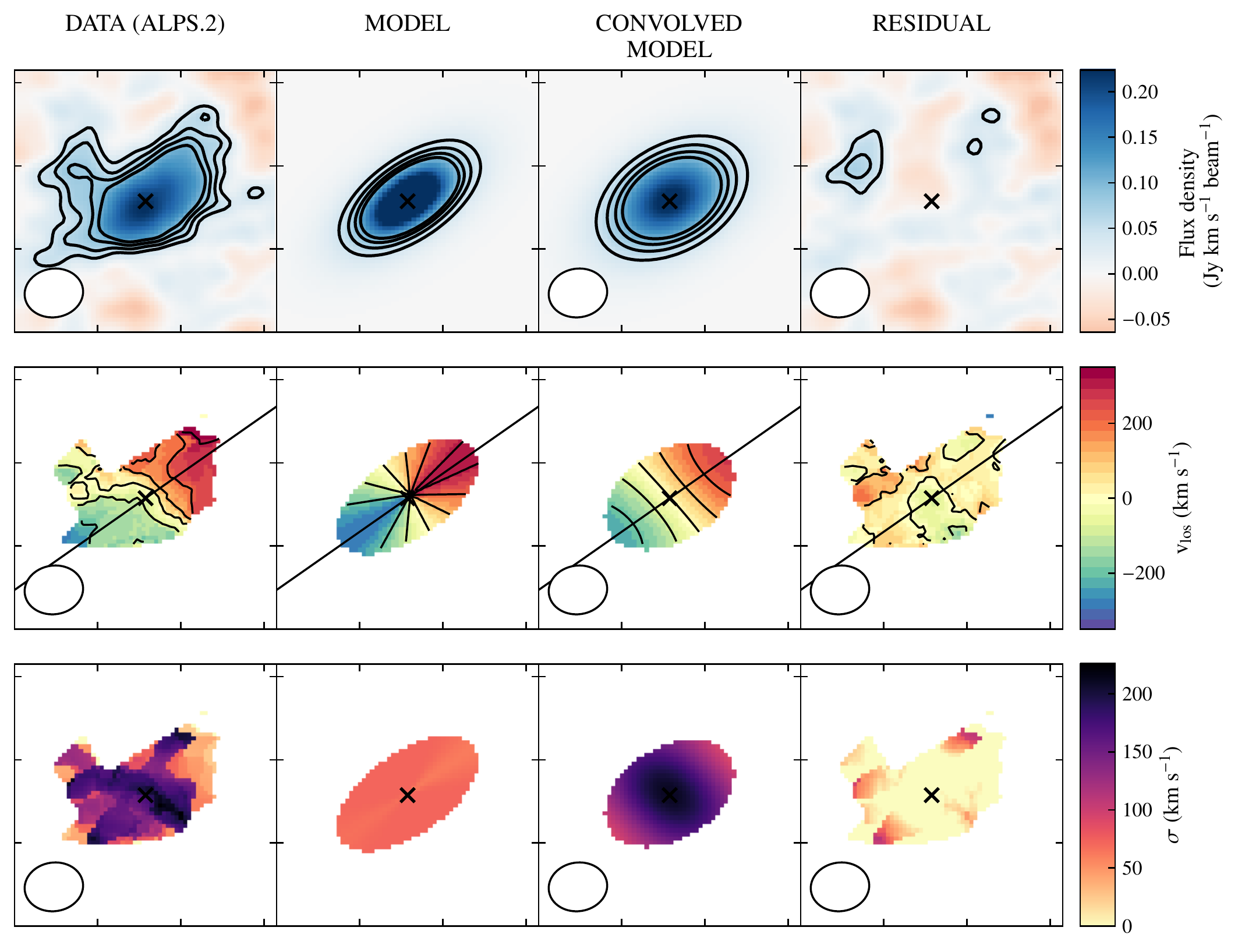}
		\end{figure*}

	\begin{figure*}[!htb]
		\ContinuedFloat
		\captionsetup{list=off}
			\centering
			\includegraphics[width=0.84\textwidth, trim={0.cm 0.cm 0.cm 0.cm},clip]{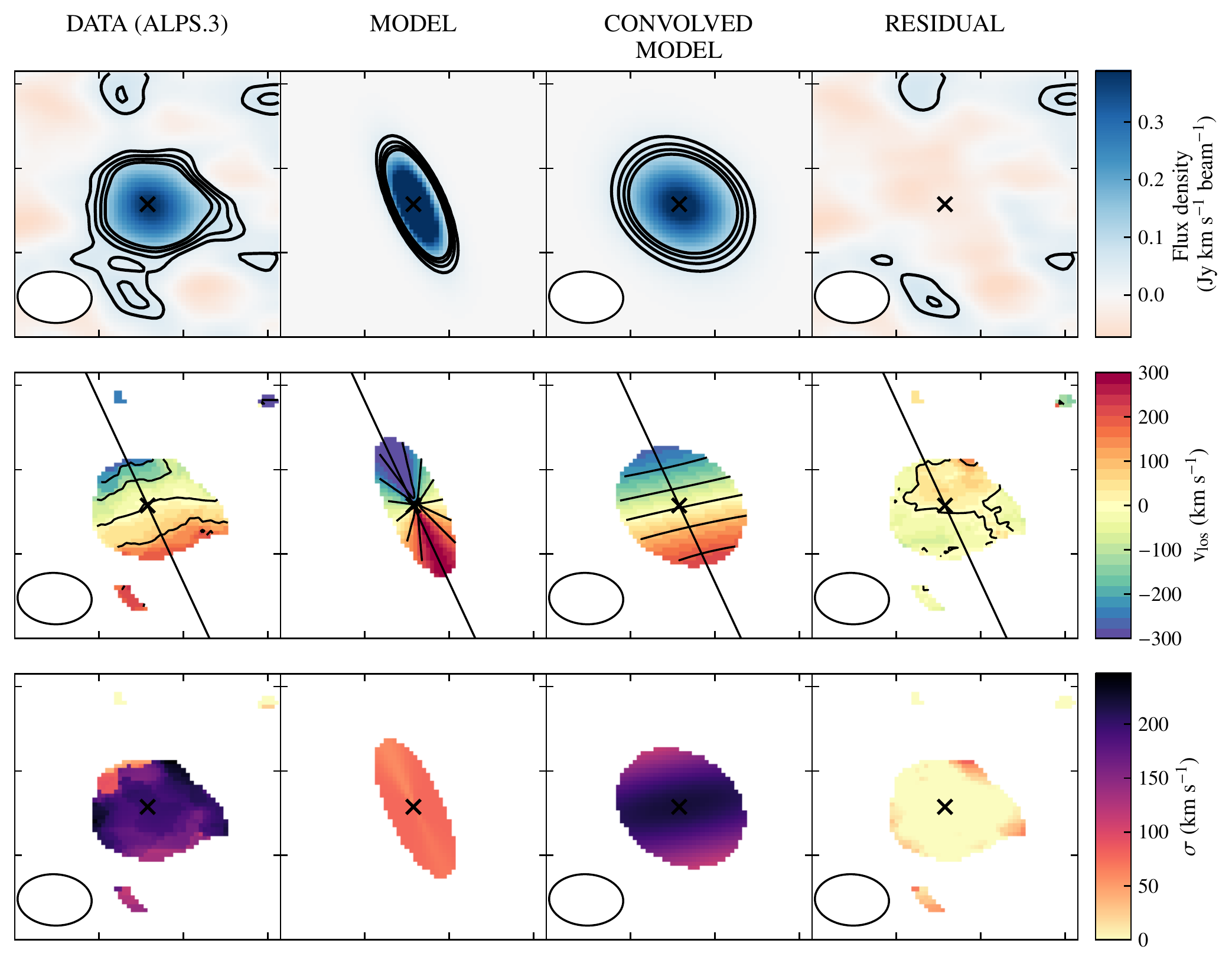}
			\caption{\texttt{Qubefit} models of the integrated CO flux (top row), velocity (middle row) and velocity dispersion (bottom row) of ALPS.1, 2 and 3 (labeled above data column for each). For ALPS.1 and 2 we observe CO(2-1) whereas for ALPS.3 we observe CO(3-2). From left to right: data, intrinsic model, convolved model, and residual (data - convolved model). The maps of the integrated velocity and velocity dispersion have been created using the $2\sigma$ blanking method (described in Section \ref{sub:kinematic_modeling}). For clarity, the maps are additionally masked at the outer $2\sigma$ contour of the integrated intensity (top row) for the respective column (data, model and convolved model). The best-fit position angles are indicated by the black lines.  \label{fig:qubefit_maps_alps}}		
	\end{figure*} 
	

	\begin{table} 
\caption{Inferred Dynamical Properties \label{tab:dynamical_props}} 
\begin{tabular}{@{}>{\raggedright}m{0.34\columnwidth}>{\centering}m{0.18\columnwidth}>{\centering}m{0.18\columnwidth}>{\centering\arraybackslash}m{0.18\columnwidth}@{}}
\toprule 
Source &ALPS.1 & ALPS.2 & ALPS.3\\ 
\midrule 
\midrule 
\multicolumn{4}{@{}l}{Kinematic Analysis with \texttt{qubefit}}\\ 
\midrule 
Inclination (degrees)& 76 $^{+10}_{-4}$ & 60 $^{+4}_{-4}$ & 69 $^{+6}_{-6}$  \\ 
$r_{1/2}^{\mathrm{CO}}$ (kpc)  & 3.7 $^{+0.3}_{-0.3}$ & 4.6 $^{+0.3}_{-0.3}$ & 2.6 $^{+0.2}_{-0.3}$  \\ 
$v_\mathrm{rot,max}$ (km s$^{-1}$) & 279 $^{+8}_{-14}$ & 355 $^{+12}_{-14}$ & 349 $^{+19}_{-23}$  \\ 
$\sigma$ (km s$^{-1}$)& 57 $^{+11}_{-12}$ & 73 $^{+13}_{-14}$ & 77 $^{+10}_{-11}$  \\ 
$v_\mathrm{rot,max}/\sigma$& 4.9 $\pm$ 1.0& 4.9 $\pm$ 0.9& 4.5 $\pm$ 0.7 \\ 
\midrule 
\midrule 
\multicolumn{4}{@{}l}{Dynamical analysis within $2\, r_\mathrm{1/2}^\mathrm{CO}$}\\ 
\midrule 
$M_\mathrm{dyn}$ ($10^{11}$ M$_\odot$) & 1.3 $\pm$ 0.2& 2.7 $\pm$ 0.3& 1.5 $\pm$ 0.2 \\ 
$M_\mathrm{baryon}$ ($10^{11}$ M$_\odot$) & 1.9 $\pm$ 0.7& 1.3 $\pm$ 0.6& 2.4 $\pm$ 0.9 \\ 
$f_\mathrm{DM}$ & $<$0.3& 0.5 $\pm$ 0.2& $<$0.4 \\ 
\midrule 
\midrule 
\multicolumn{4}{@{}l}{Dynamical analysis within $6\, r_\mathrm{d}^\mathrm{CO}$}\\ 
\midrule 
$M_\mathrm{dyn}$ ($10^{11}$ M$_\odot$) & 2.4 $\pm$ 0.3& 4.8 $\pm$ 0.5& 2.6 $\pm$ 0.4 \\ 
$M_\mathrm{baryon}$ ($10^{11}$ M$_\odot$) & 2.6 $\pm$ 1.0& 1.7 $\pm$ 0.9& 3.6 $\pm$ 1.3 \\ 
$f_\mathrm{DM}$ & $<$0.2& 0.6 $\pm$ 0.2& $<$0.3 \\ 
\midrule 
\bottomrule 
\end{tabular} 
\begin{itemize}[leftmargin=*] 
{\item[$^*$] The baryonic masses and hence dark matter fractions depend on the inferred $\alpha_\mathrm{CO}$, provided in Table \ref{tab:derived_prop}}. 
 {\item[$^{**}$] The quoted uncertainties are at the 1$\sigma$ level whereas the upper limits are at 2$\sigma$.} 
 \end{itemize} 
\end{table}

	\subsection{Kinematic modeling} 
		\label{sub:kinematic_modeling}

		We constrain the rotation velocity and velocity dispersion of each of the three galaxies using the python-based kinematic modeling tool {\sc{Qubefit}}, \cite[described in Appendix C. of][]{2019ApJ...882...10N}\footnote{\url{https://github.com/mneeleman/qubefit}}. {\sc{Qubefit}} fits a model cube to the data, convolving the model to the same spectral and spatial resolution as the data. The model is compared to the data via the $\chi^2$ statistic, and takes into account the spatial correlation between pixels via a bootstrapping analysis. The full parameter space is sampled via the Markov Chain Monte Carlo method, thereby yielding likelihood distributions for the parameters being fit. 

		We model the line emission as stemming from a thin, exponential disk with only circular velocities in the plane of the disk, i.e. we do not consider any radial motions. We fit for the source centre, maximum rotational velocity, $v_\mathrm{rot,max}$, velocity dispersion, $\sigma_v$ and the scale length of the exponential disk, $r_d$. The median and $\pm 1\sigma$ values, for each galaxy, are provided in Table \ref{tab:dynamical_props}. We choose not to place a prior on the inclination or PA as the CO emission may probe a different region than the rest-frame optical data. For the initial guesses of the inclination and position angle, we use the best-fit values of \cite{2012ApJS..203...24V}, scaling the best-fit axis ratio, $q$, to an inclination via $\cos ^2 i = q^2$. We fit the model cube to regions of 3$''$ diameter (larger regions resulted in poorer constraints). We note that the best-fit rotation velocities and inclination angles are highly degenerate. However, for all three galaxies we recover inclination angles consistent with those inferred based on the axis ratios fit to the \emph{HST} data. For ALPS.3, the best-fit position angle differs by 20$^\circ$ from the position angle fit to the \emph{HST} data. Forcing the position angle to match decreases the rotation velocity to a value 20 \kms\, lower than that quoted in Table~\ref{tab:dynamical_props}. We caution that the inferred velocity dispersions are global estimates that include dispersions due to motion along the line-of-sight (i.e. due to motion inside a thick disk, or, motions due to warps) as well the true velocity dispersion of the gas. Thus, the dispersion inferred via {\sc{Qubefit}} is an upper limit on the intrinsic gas velocity dispersion. Although we provide the estimated errors for the thin disk model in Table~\ref{tab:dynamical_props}, we treat the dispersion values as upper limits. The CO half-light radii measured with {\sc{QUBEFIT}} are lower than measured from the moment-0 maps with {\sc{GALFIT}}, by at least $20\%$. We conclude that this is mainly due to the low S/N of the emission in individual 50 \kms\ channels compared to that in the map collapsed over 800 \kms.

		We compare the model moment maps to those of the data in Figure~\ref{fig:qubefit_maps_alps}. To create the moment-1 and 2 maps for the data, we implement a {\sc python}-based algorithm that identifies coherant source emission, associated with a $>3\sigma$ peak in both the spatial and spectral axis \citep[see also][]{2009AJ....137.4670L}. To do this, we expand the region around each $>3\sigma$ peak outwards, in RA, DEC and velocity, until we reach a $2\sigma$ boundary. All pixels outside these boundaries are masked when making the moment-1 (intensity-weighted velocity) and moment-2 (intensity-weighted velocity dispersion) maps. This method produces moment maps that are more representative of the velocity fields than if we include all pixels, but some artifacts, such as the $400\,\kms$ components for ALPS.3, still remain. The residuals of the moment 0 - 2 maps indicate a good quality of fit (see Figure~\ref{fig:qubefit_maps_alps}) and the rotation velocity and dispersion values are consistent with the best-fit parameters inferred using the 3D modeling algorithm {\sc{3DBarolo}} \citep{2015MNRAS.451.3021D}. However, for any fitting method the degeneracy between the rotation velocity and inclination is large.
		
		Based on the data presented here, we note that single-Gaussian line profiles can severly overestimate the disk rotation. From the single Gaussian fit to the CO spectrum (of the ASPECS LP data), presented in \cite{2019ApJ...882..139G}, the FWHM of the CO(2-1) line for ALPS.2 (ASPECS LP 3mm.05) is $\mathrm{FWHM}=620\pm 60\, \kms$. If the gas exhibits ordered motion then the maximum rotation velocity can be approximated by $v_\mathrm{rot} = 0.75 \, \mathrm{FWHM}/\sin i $ \citep[see e.g.][]{2013ApJ...773...44W,2015ApJ...801..123W,2018ApJ...854...97D}. For ALPS.2, this results in an estimated rotation velocity of $\sim 510\, \pm\, 50 \, \kms$, significantly larger than the value measured from the 3D modeling.


	\subsection{Dynamical Masses} 
		\label{sub:dynamical_masses}

		Using the inferred kinematic properties, we estimate the total amount of matter within the region probed by the CO line emission, i.e. the dynamical mass, $M_\mathrm{dyn}$. For local galaxies, the dynamical mass is typically estimated from the circular motion, inferred from a rotation curve, and the maximum extent of the rotating disk, typically inferred from HI or stellar light profiles. In combination with measurements of the baryonic mass components (i.e. the stellar and gas mass), the dynamical mass can be applied to estimate the fraction of baryonic or dark matter within the observed region of the galaxy. Such estimates become highly uncertain for high-redshift galaxies, for which high-resolution observations of line kinematics are scarce, stellar and gas masses are highly uncertain, source sizes are challenging to constrain and the inclination is often unknown. Moreover, even for galaxies for which the emission line observations are resolved, the observations typically do not extend to the radii required to infer the dynamical mass within the entire baryonic disk. 

		We infer the dynamical mass of all three galaxies, assuming that the bulk of the emission can be well described by a rotating disk (which appears consistent with the lines profiles, as discussed in Section~\ref{sub:line_kinematics}). The correct choice of radius is somewhat arbitrary (often high-redshift studies use the rest-frame optical half-light radius or twice this value to estimate dynamical masses). Here, we estimate these values for two definitions of the maximum radius probed by the CO, twice the half-light radius $r_\mathrm{max} = 2 r_\mathrm{1/2}^\mathrm{CO}$ and six times the exponential scale length $r_\mathrm{max} = 6 r_\mathrm{d}^\mathrm{CO}$. For consistency with the inferred maximum rotation velocities, we estimate the outer radii using the half-light radii inferred using {\sc{QUBEFIT}} (Table \ref{tab:dynamical_props}). 

		We determine the dynamical mass and total baryonic mass (stellar plus molecular gas mass) from within $r_\mathrm{max}$, via,
		\begin{align}
			M_\mathrm{dyn} (\leq r_\mathrm{max}) = \dfrac{ v_\mathrm{rot}^2 r_\mathrm{max} }{G}\, ,
		\end{align}
		where $v_\mathrm{rot}$ is the modeled, rotation velocity and $G$ is the gravitational constant. We infer the total content of baryonic matter using the measured molecular gas and stellar masses in combination with the modeled surface brightness profiles of the CO and 1.6\,\textmu m emission, respectively. We estimate the fraction of light within $r_\mathrm{max}$, from the 2D model profiles, and scale the molecular gas and stellar masses by the respective fractions. We thereby assume that the 1.6\,\textmu m emission perfectly traces the stellar mass, and, that the excitation and CO-to-molecular gas mass conversion are constant over the entire galaxy disk.  The dynamical and baryonic masses estimated for the three sources are provided in Table \ref{tab:dynamical_props}. For ALPS.2, we estimate a dynamical mass of $(2.7\,\pm\, 0.3) \times 10^{11}\Msun$ within $2 r_\mathrm{1/2}^\mathrm{CO} \approx 9.2$\, kpc, very close to the $M_\mathrm{dyn} \sim 2.0 \times 10^{11}\, \Msun$ inferred by \cite{2016ApJ...833...70D} (from the initial 3\,mm ASPECS data, based on the rest-frame optical half-light radius of 8.3 kpc). 

		As described in Section \ref{sub:stellar_masses}, a number of assumptions affect the inferred stellar masses. We therefore account for a factor of two uncertainty on the stellar mass (see Appendix \ref{sec:sed_analysis}). The molecular gas masses are based on the assumption that the mass-metallicity relation holds for all three galaxies and that the $\aco$ conversion factor is purely a function of the metallicity and doesn't evolve with redshift. We include the uncertainties on the metallicity estimate and $\aco$ based on the uncertainties of the empirical relations used. We attempt to account for the uncertainty on the CO excitation. For ALPS.3 we use the measured line luminosity ratio, whereas for ALPS.1 and 2 we assume a CO(2-1)-to-CO(1-0) line ratio consistent with what has been previously measured for high-redshift sources. In addition, we assume that the ratio of $\Mstar/L_\mathrm{1.6\mathrm{\mu} m}$ is constant, and, that the line ratios and $\aco$ are constant over the entire galaxy disk. Compared to other systematic uncertainties, radial variations in the applied ratios will have little impact on the derived baryonic masses and dark matter fractions. In addition to these uncertainties on the baryonic mass, the dynamical mass is heavily dependent on the inclination angle. We account for the range of inclination angles that fit the CO data in the uncertainties on the maximum rotation velocities.

		Based on the dynamical and baryonic mass estimates, we estimate the dark matter fractions, $f_\mathrm{DM} = 1- M_\mathrm{baryon}/M_\mathrm{dyn}$, enclosed within the two types of maximum radii chosen here. For ALPS.2, we infer dark matter fractions of up to 70\% and 80\% (within $1\sigma$), within $r_\mathrm{1/2}^\mathrm{CO}$ and $6 r_\mathrm{d}^\mathrm{CO}$, respectively. We place $2\sigma$ upper limits on the dark matter fractions of ALPS.1 and 3 as the measured values are consistent with 0.
	


	\begin{figure*}
		\centering
		\includegraphics[width=\textwidth, trim={1.3cm 0.3cm 1.3cm 0.5cm},clip]{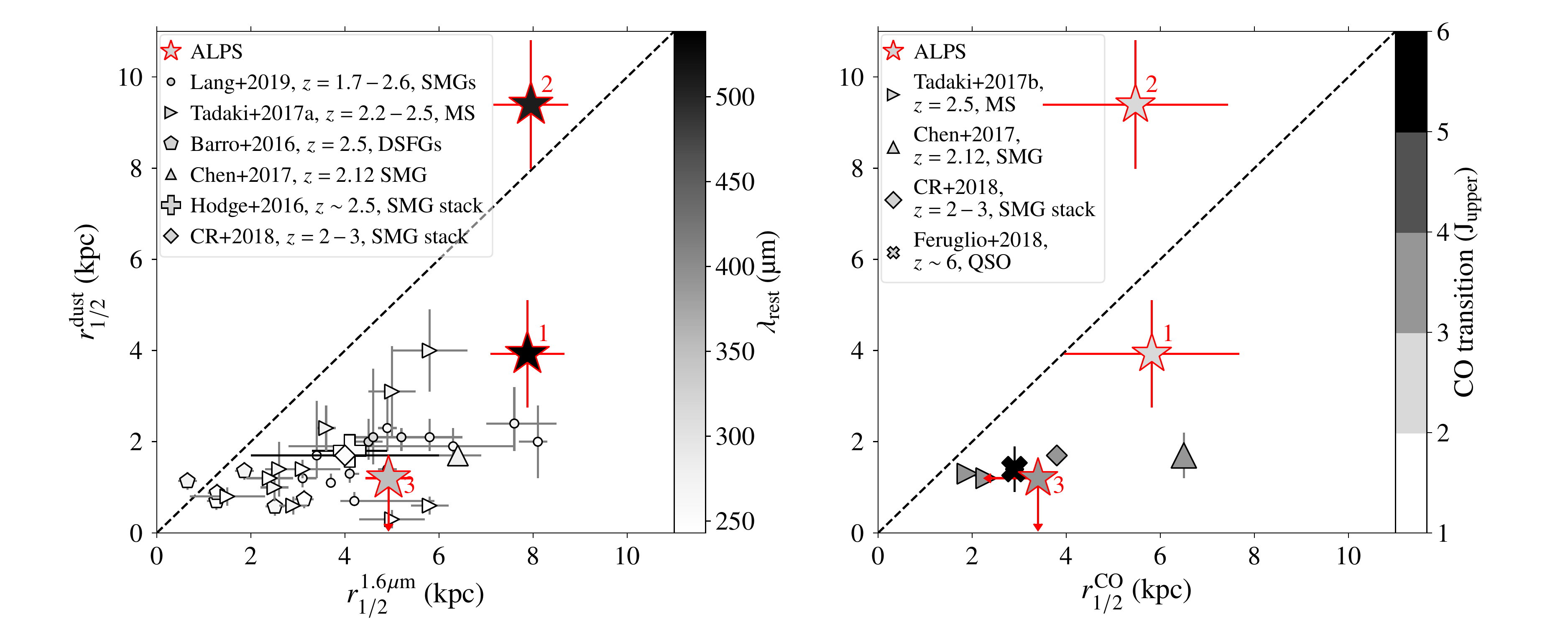}
		\caption{ Comparison of the source sizes measured for our sample (stars) to other samples with dust-continuum size measurements \citep[labeled in legends][]{2016ApJ...827L..32B,2016ApJ...833..103H,2017ApJ...846..108C,2017ApJ...834..135T,2017ApJ...841L..25T,2018ApJ...863...56C,2018A&A...619A..39F,2019ApJ...879...54L}. The source sizes shown here are measured from 2D maps using {\sc{GALFIT}} (note that for the CO emission we also estimate sizes based on the 3D data as described in Section \ref{sub:kinematic_modeling}). Left panel: the dust half-light radii compared to the half-light radii measured from 1.6\,\textmu m emission. The greyscale of the points indicates the wavelength of the rest-frame dust-continuum emission. For most high-redshift galaxies observed to date, the dust continuum is more compact than the 1.6\,\textmu m emission. ALPS.2 is rare in that we measure equivalent dust and stellar half-light radii.  Right panel: the dust half-light radii compared to the CO half-light radii. Here the greyscale indicates the upper level of the observed CO transition. As highlighted in this panel, very few high-redshift sources have the high-resolution observations of both CO and dust-continuum emission, needed to compare the relative spatial distributions. \label{fig:size_comp}}
	\end{figure*}

\section{Discussion} 
	\label{sec:discussion}

		\subsection{Comparing the Spatial Extent of CO, Dust and Stellar Emission} 
			\label{sub:comparing_the_spatial_extents_of_co_dust_and_stellar_emission}

			We have measured the spatial extent of the 1.3\,mm dust-continuum, CO line and rest-frame optical emission of three star-forming galaxies at $z=1.41, 1.55$ and $2.70$. In this section, we discuss our findings and compare our results to simulations as well as previous observational studies of local and high-redshift galaxies. Although there is a wealth of high-resolution data available for local galaxies, few resolved observations exist of high-redshift sources that are not gravitationally lensed (summarised in Figure \ref{fig:size_comp}). 

			\subsubsection{Limitations}

				We begin by noting that our data impose some limitations upon the methodology applied here. To estimate galaxy sizes, we fit a single disk component to all sets of data, assuming that the inferred half-light radius is representative of the entire disk. We do so because the resolution ($\sim 6$ kpc for the CO) and sensitivity of the ALMA observations does not allow us to decompose the data into multiple components. However, stellar bulges are known to produce a steep, central rise for the rest-frame optical emission, and high central SFRs do the same for dust-continuum emission. Thus, local studies typically either decompose the stellar and continuum emission into a bulge and disk component, or omit the central few kpc when characterising the scale length of the exponential disk \citep[e.g.][]{2009ApJ...701.1965M,2015A&A...576A..33H,2017A&A...605A..18C}.


			\subsubsection{CO vs Rest-frame Optical Sizes}

				For all three galaxies, the CO half-light radii are at least $30\%$ smaller than the rest-frame optical half-light radii (see Table \ref{tab:source_sizes}). This difference appears to be slightly larger than what has been observed for most local and high-redshift galaxies. As for our sources, \cite{2017A&A...605A..18C} find that the mean scale length of the CO(2-1) and CO(1-0) surface brightness profiles for their 18, face-on local galaxies is, $20-60\%$ smaller ($\sim 40\%$, on average) than the mean scale length at 0.6\,\textmu m (equivalent to the observed-frame 1.6\,\textmu m for our $z\sim 1.5$ galaxies). However, for the local galaxies in the HERA CO line Extragalactic Survey (HERACLES \citealt{2009AJ....137.4670L}) and SINGS samples \citep{2006ApJ...652.1112R} the measured CO scale lengths are, on average, consistent with the scale lengths of the 3.6\,\textmu m emission, but range from half to almost twice the scale length of the stellar tracers. Similarly, recent studies of $z=2-3$ SMGs find equivalent CO and rest-frame optical half-light radii \citep{2017ApJ...846..108C,2018ApJ...863...56C}. 

				The relative size of the CO versus rest-frame optical emission depends on the observed CO transition and rest-frame wavelength.Results from the Physics at High Angular resolution in Nearby GalaxieS \citep[PHANGS,][]{2019Msngr.177...36S} Survey as well as from HERACLES \citep{2009AJ....137.4670L} indicate that the CO(2-1) emission is co-spatial with the CO(1-0) emission. However, the ratio of the line luminosities can vary between 0.6-1.0, with brighter CO(2-1) typically found in the centers of galaxies. Brighter, or more peaked, CO emission may therefore bias our inferred measurements to smaller radii, particularly because our observations are not sensitive to gas columns below $\sim 5\times 10^{21} \mathrm{cm}^{-2}\, \mathrm{beam}^{-1}$. The comparison between CO and stellar disk sizes is also influenced by the observed wavelength of the stellar emission. Age gradients in the stellar population across the disk and/or increased effects of reddening in the central regions, also lead to a decline in the measured radii from the UV to NIR \citep[e.g.][]{1994A&AS..108..621P,1996A&A...313..377D,1997A&AS..124..129P,2003ApJ...582..689M,2010MNRAS.406.1595F,2006A&A...456..941M}.\footnote{We also find that the stellar half-light radii of our galaxies decrease by up to 20\% from the observed-frame 1.25\,\textmu m to 1.6\,\textmu m emission, e.g. the 1.25\,\textmu m half-light radius of ALPS.2 is $9.8\pm 0.9$ kpc, which is $20\pm 10\%$ larger than the 1.6\,\textmu m size.} Thus, the size measured for the 1.6\,\textmu m emission is also likely to be smaller than the intrinsic size of the stellar disk. 

			\subsubsection{Dust Continuum vs Rest-frame Optical Sizes}

				We observe both compact and extended dust emission, amongst the three galaxies studied here.
				For ALPS.1 and 3, the 1.3\,mm half-light radii are $(50 \pm 20) \%$ and $<70\%$ of the 1.6\,\textmu m half-light radii, respectively. In contrast, the 1.3\,mm half-light radius of ALPS.2 is equivalent to that at 1.6\,\textmu m.

				Like the rest-frame 350\,\textmu m emission studied here, the rest-frame 240\,\textmu m (observed-frame 870\,\textmu m) dust-continuum emission of ALPS.3 is very compact. \cite{2019ApJ...882..107R} analyse this emission at $0\as03$ resolution. By modeling the source emission in the $uv$-plane, using two concentric, elliptical Gaussians, they find that the ``extended'' dust component is best described by a profile with a major axis FWHM of $\sim 3.6 \pm 0.1$ kpc, corresponding to a half-light radius of $\sim 1.8$ kpc, similar to the half-light radius measured here for the rest-frame 350\,\textmu m emission (i.e. $<1.2$\,kpc from the image-plane and $\sim 1.2 \pm 0.2$ kpc from the $uv$-plane analysis). Assuming that the continuum emission can be modeled by a modified blackbody with a temperature of 25 K and dust emissivity index of 1.8, the rest-frame 350 \textmu m ASPECS LP measurement, $S_\mathrm{\nu} = 1070 \pm 50$\,\textmu Jy, extrapolates to a rest-frame 240\,\textmu m flux of $\sim 2890 \pm 120$\,\textmu Jy. This is consistent with the $2780 \pm 90$\,\textmu Jy measured by \cite{2019ApJ...882..107R}, implying that the shorter wavelength observations have not missed a significant component of extended emission. Thus, the shorter wavelength data is further evidence that the dust component of ALPS.3 is very compact.

				The spatial extent of the dust-continuum emission in ALPS.1 and 2 differs significantly, despite the two galaxies exhibiting similar rest-frame optical half-light radii and similar stellar masses and SFRs (Table \ref{tab:derived_prop}). From the multi-wavelength comparison in Figure~\ref{fig:thumbnails}, it would appear as though the dust-continuum emission traces the bright, central stellar component visible in the 0.6 and 1.6\,\textmu m images (second and third columns from the left) for ALPS.1, with no 1.3\,mm emission above our detection limit visible for the southern tail of the optical emission. Conversely, the dust emission from ALPS.2 appears to coincide with the location of the clump-like, outer disk components visible in the 0.6 and 1.6\,\textmu m images (columns 2 and 3 of Figure~\ref{fig:thumbnails}), but not the central, red, stellar component. We also note that an additional blue clump is apparent to the north of the UV-NIR images of ALPS.2, which is not apparent in dust-continuum emission, but it is unclear if this is a fore- or background source.

				The extended dust-continuum emission of ALPS.2 differs not only from what is observed for ALPS.1 and 3, but also from other high-redshift galaxies (Figure \ref{fig:size_comp}). In most of these, the dust emission is $2-4\times$ more compact than the stellar emission, e.g. for SMGs the 870\, \textmu m half-light radii are typically measured to be $\lesssim 3$ kpc, whereas the 1.6\,\textmu m emission can extend to $8$ kpc \citep{2015ApJ...799...81S,2015ApJ...798L..18H,2019ApJ...876..130H,2017ApJ...846..108C,2018ApJ...863...56C,2019ApJ...879...54L,2019MNRAS.490.4956G}. However, based on observations and radiative transfer models of local galaxies, extended dust disks with sizes comparable to the size of the stellar disk are the norm \citep{1999A&A...344..868X,2007A&A...471..765B,2009ApJ...701.1965M,2012A&A...541L...5H}. This tension can be resolved by the fact that the dust-continuum emission does not perfectly trace the bulk of the dust mass, but may be temperature-weighted and thus also sensitive to the sources of heating.

				The dust-continuum emission may be very compact, if the central regions are dominated by starbursts (or AGN). Based on their FIRE simulations of high-redshift SMGs, \cite{2019MNRAS.488.1779C} find that the FIR continuum emission mostly stems from regions with high star formation rates, tracing both the central region as well as spiral arms. Similarly, based on Illustris TNG simulations of $1<z<3$ MS galaxies, \cite{Popping_prep}, find that the $1.2$\,mm emission best traces the regions with high rates of star formation. Moreover, both studies find that the $\sim 1$\,mm dust half-light radii are typically smaller than the dust half-mass radii. These studies indicate that the centrally-concentrated continuum emission observed for many high-redshift galaxies has traced the high star formation rates of the central regions. 

			\begin{figure*}
				\centering
				\includegraphics[width=0.49\textwidth, trim={1.cm 0.6cm 0.cm 0.cm},clip]{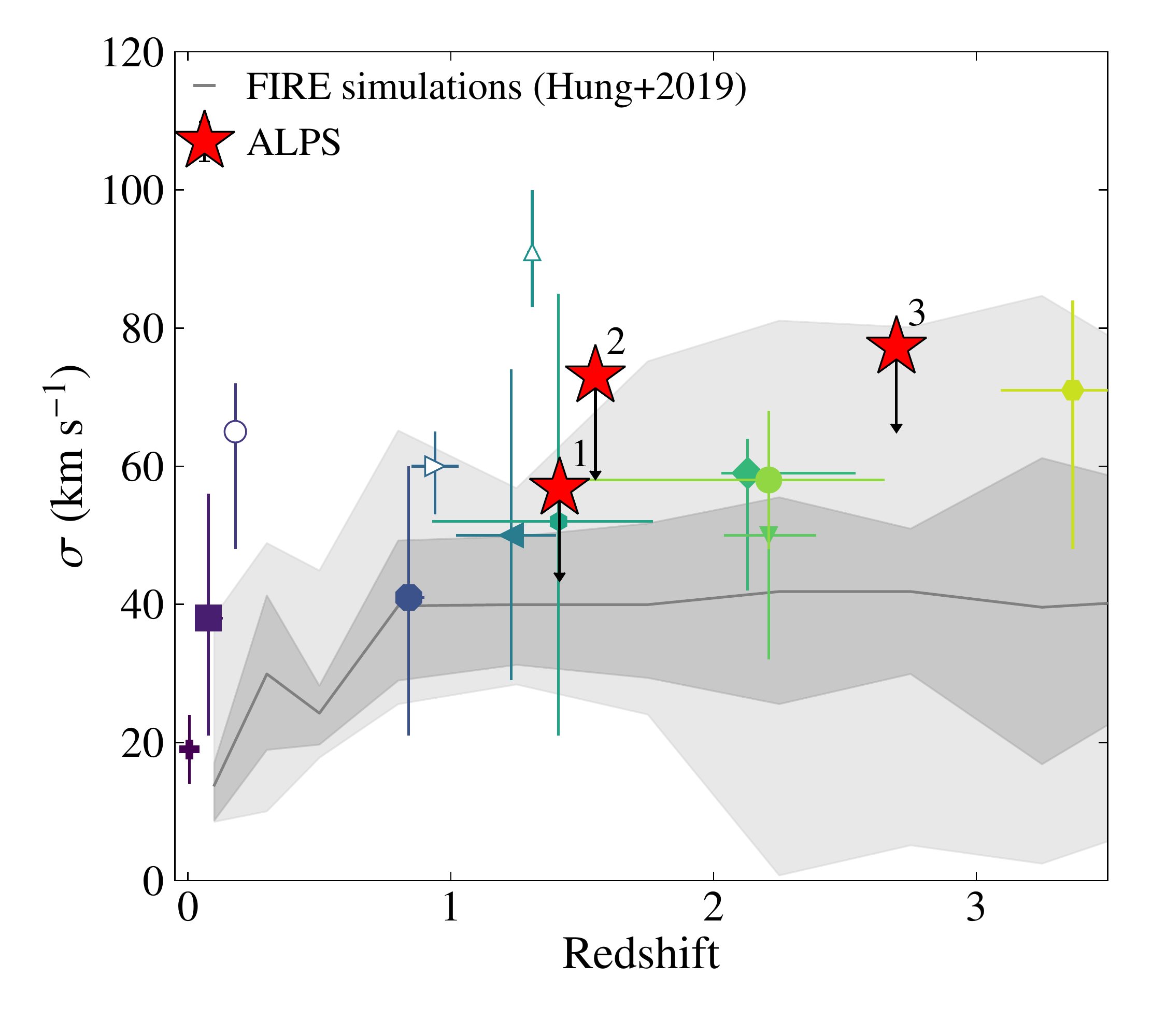}
				~
				\includegraphics[width=0.49\textwidth, trim={1.cm 0.6cm 0.cm 0.cm},clip]{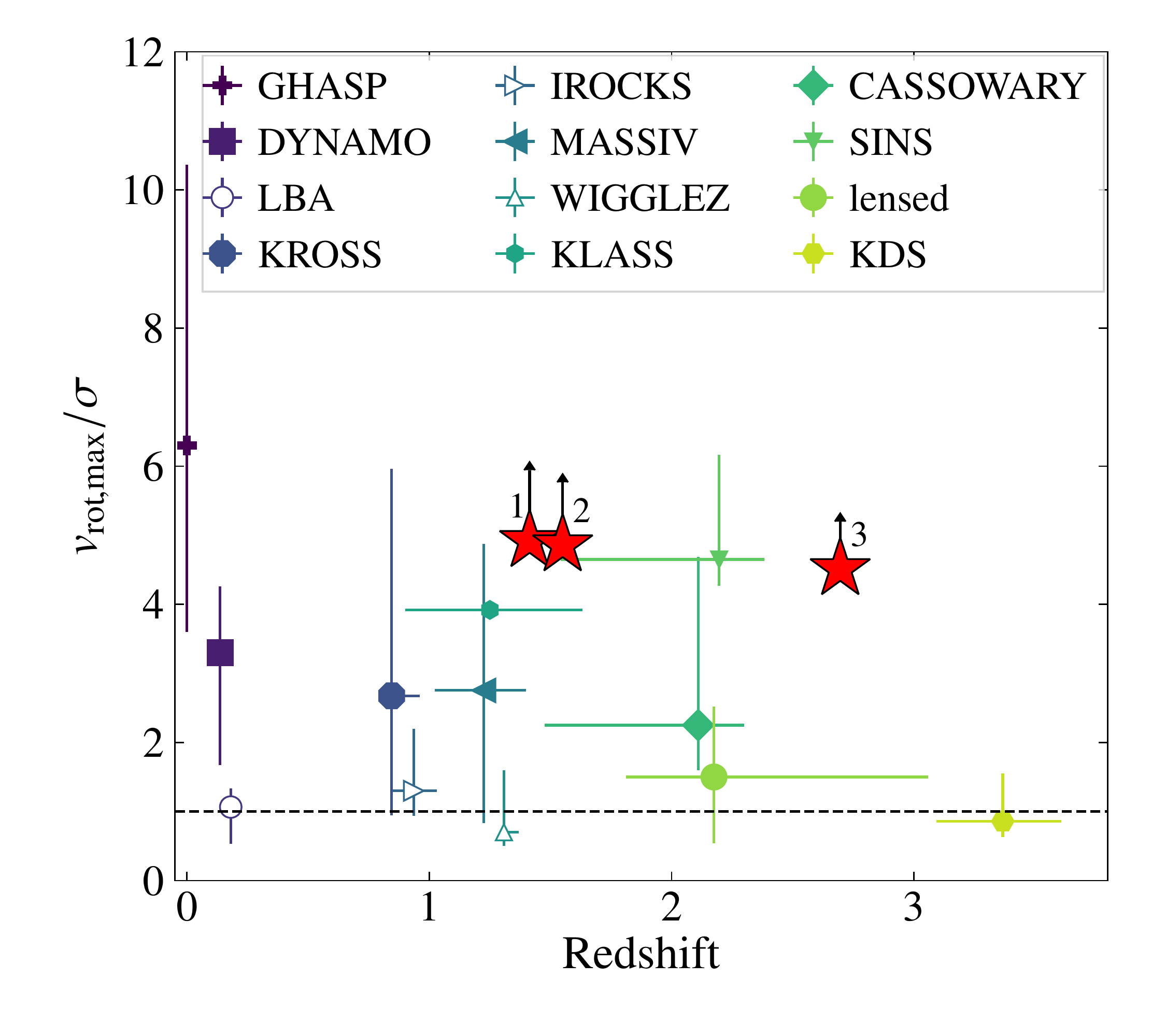}
				\caption{Left: Velocity dispersion as a function of redshift for star-forming galaxies. The solid grey line shows the median value for the simulated MS galaxies of \cite{2019MNRAS.482.5125H} with the dark and light grey regions enclosing 68\% and 95\% per cent, respectively, of the simulated sources. The velocity dispersions of ALPS.1, 2 and 3, denoted by the red stars, are consistent with previous observations of MS galaxies. Right: comparison of the ratio of rotation-to-random motions for other galaxy samples. The colored symbols and errorbars in both panels indicate the median and 18th-64th percentiles of the GHASP survey \citep{2010MNRAS.401.2113E}, DYNAMO survey \citep{2014MNRAS.437.1070G,2019MNRAS.485.5700S}, Lyman Break Analogues (LBA) \cite{2010ApJ...724.1373G}, KROSS \citep{2017MNRAS.467.1965H,2018MNRAS.474.5076J}, MASSIV \citep{2012A&A...539A..92E}, WiggleZ survey \cite{2011MNRAS.417.2601W}, KLASS \citep{2017ApJ...838...14M}, CASSOWARY \citep{2016ApJ...820...84L}, SINS \citep{2009ApJ...697..115C}, Lensed galaxies \citep{2015MNRAS.450.1812L} and KDS \citep{2017MNRAS.471.1280T}. The ALPS galaxies studied here have higher ratios of rotation-to-random motion than the majority of high-redshift galaxies observed to date. The velocity dispersion values for the comparison samples are provided in Table 2 of \citealt{2019MNRAS.482.5125H}. Observations for which the beam smearing is taken into account (for the rotation velocity) are shown as filled symbols (see Appendix A of \citealt{2017MNRAS.471.1280T}).  \label{fig:vdisp}}
			\end{figure*}

			\subsubsection{CO vs Dust Continuum}

				Like the comparison to the rest-frame optical emission in the previous section, we find that the dust-continuum emission is more compact than the CO line emission from ALPS.1 and 3, but more extended than the CO emission from ALPS.2. The compact continuum emission, relative to the CO emission, appears consistent with previous observations of unlensed high-redshift galaxies \citep{2017ApJ...841L..25T,2017ApJ...846..108C,2018ApJ...863...56C,2018A&A...619A..39F}, as summarised in Figure \ref{fig:size_comp}. 

				Despite the relative differences in the extent of the CO and dust-continuum emission, the molecular gas masses inferred based on the total dust-continuum flux densities, using the empirical calibration of \cite{2014ApJ...783...84S}, are consistent with those measured based on the total CO emission. Thus, even though the total dust-continuum luminosity may correlate well with the CO emission (yielding the same total gas mass estimates), the dust-continuum emission is not a straightforward tracer of the extended dust disk \citep[see also][]{2015ApJ...799...96G}. 

				In addition to the impact of dust heating, other physical effects may contribute to the difference in the extent of the CO and dust-continuum emission. Significant increases in the gas-to-dust ratio (GDR) towards the outskirts of galaxies may result in measured size differences. Although some individual galaxies exhibit evidence for an increase in the GDR with radius \citep[e.g.][]{2010MNRAS.402.1409B} for most local galaxies the measured variation appears negligible \citep{2013ApJ...777....5S}. Differences in the temperature and opacity across the disk seem likely to play a larger role, as indicated by the comparison of radiative transfer models with both the stacked CO(3-2) and dust-continuum emission in \cite{2018ApJ...863...56C}. We note that the dust-continuum emission of ALPS.1 and 3 appear steeper than for ALPS.2, indicating more centralised heating in these galaxies. Although the X-ray observations and SED strongly imply the presence of a central AGN in ALPS.2, there is no evidence from the UV-to-NIR images or 1.3\,mm dust continuum for increased heating in the centre. Deeper, high resolution continuum observations will help to shed light on this issue.

		\subsubsection{A Plausible Explanation for Compact Dust and CO Emission} 
			\label{sub:mock_high_redshift_starbursts}

				The compact dust (and CO) observations for ALPS.1 and 3 are consistent with what would be observed for galaxies hosting nuclear starbursts. To illustrate this, we use two local galaxies hosting nuclear starbursts, NGC 253 and NGC 4945, as an example (see Appendix \ref{sec:local_starbursts}). Like the galaxies studied here, both of these local galaxies are heavily inclined and extended (with the 870\,\textmu m emission observed out to a radial distance of $\sim 7.5$ kpc). Using the 870\,\textmu m LABOCA observations of these two galaxies \citep{2008A&A...490...77W}, we create mock 1.3\,mm maps with the same total flux density, resolution and S/N as the data for ALPS.1, 2 and 3 presented here. 

				We find that the convolved size at half the peak surface brightness of the mock 1.3\,mm maps (blue contours), matches the beam of our 1.3\,mm data (Figure \ref{fig:local_starbursts}), due to the high contrast between the central, dust-continuum emission and that of the disk (up to a factor of 50 for NGC 4959).  With the S/N and resolution of the data for ALPS.1 and 2, the spiral arms of NGC 253 would not be detected (at $>3\sigma$) and only the brightest spiral of NGC4959 would be detected, making the observations appear asymmetric (similar to the 1.3\,mm maps of ALPS.1 and 2 in Figure \ref{fig:thumbnails}). The fact that we observe no extended structure in the 1.3\,mm emission of ALPS.3 implies that the disk is less extended and/or the contrast between the nuclear and disk regions is even greater than for these two local starbursts. The comparison to these simple mock observations is consistent with the fact that our 1.3\,mm observations are sensitive to a gas column of $N_\mathrm{H_2}\sim 10^{22}\, \mathrm{cm}^{-2}$, whereas the column densities measured for most of the spiral arms regions of NGC 253 are significantly lower. 
		


		\subsection{Dynamical Properties} 
			\label{sub:dynamical_properties}

			We have measured the dynamical properties of all three sources, and conclude that the gas motion of each disk is dominated by rotation. The measured velocity dispersions are consistent with the measurements from the KLASS, CASSOWARY, KROSS and SINS surveys of galaxies at $1<z<3$ \citep{2017ApJ...838...14M,2016ApJ...820...84L,2018MNRAS.474.5076J,2009ApJ...697..115C}  and the simulations of \cite{2019MNRAS.482.5125H} (Figure~\ref{fig:vdisp}). However, the maximum rotation velocities are greater than the majority of high-redshift galaxies in these samples. Thus, the rotation-to-random motions measured for our sample are higher than what has been been measured for most other high-redshift, star-forming galaxies (Figure \ref{fig:vdisp}) but consistent with the range of values typical of local disks \citep{2010MNRAS.401.2113E}. It is unsurprising that the three galaxies studied here appear more rotation-dominated than most high-redshift galaxies studied to date, because our selection of extended and inclined sources favours ``thin'' rotating disks. 

			We note that ALPS.2 was also observed as part of the SINS sample \citep[labeled GMASS 1084 in][]{2009ApJ...706.1364F}. Based on their H$\alpha$ observations \cite{2009ApJ...706.1364F} measure a rotation to dispersion ratio of $v_\mathrm{rot}/\sigma = 4.4\pm 2.1$, consistent with the $v_\mathrm{rot}/\sigma = 4.9\pm 0.9$ derived here. However, using the intrinsic, inclination-corrected rotation curve of the best-fitting model disk they infer a lower circular velocity of $v_d= 230 \pm 38$\,\kms. This lower velocity, compared to the $355 \pm 14 \, \kms$ measured here, is consistent with the smaller half-light radius of the H$\alpha$ emission compared to that of CO ($3.1\pm 1$ kpc vs $4.6\pm0.3$ kpc), assuming that the rotation curve is still rising.

			To compare to the literature, we inferred the dark matter fractions of our sample from within two sets of outer radii, i.e. out to twice the CO half-light radii, $2 r_\mathrm{1/2}$, and to six times the CO exponential scale length, $6 r_\mathrm{d}$. For ALPS.1 and 3 the dynamical mass is of the same order as the baryonic mass, suggesting that the baryonic matter dominates at small radii. Even supposing that the molecular-to-gas mass conversion is overestimated by a factor of $\sim 4$ this would still be the case. In contrast, for ALPS.2, the measured dark matter fraction within $2 r_\mathrm{1/2}^\mathrm{CO(2-1)}$ is consistent with most simulations and local observations \citep{2016AJ....152..157L,2018MNRAS.481.1950L,2019MNRAS.489.5483T}. It is also consistent with the dark matter fractions inferred from the high-redshift observations of \cite{2019MNRAS.487.4856M} and the $0.6<z<2.2$ galaxies of \cite{2019MNRAS.485..934T}. The smaller inferred dark matter fractions of ALPS.1 and 3, compared to ALPS.2, are likely related to the more centrally-concentrated molecular gas and stellar components. It is possible that these data simply do not extend far enough to measure the maximum radial velocities. 
	


\section{Summary} 
	\label{sec:conclusions}

	We have compared ALMA and \ac{hst} observations, tracing the stellar, dust and molecular gas components of three, star-forming galaxies, ALPS.1 at $z = 1.41$, ALPS.2 at $z = 1.55$ and ALPS.3 at $z=2.70$. These galaxies were selected from the ASPECS LP, based on their bright CO and dust-continuum emission and their large, rest-frame optical sizes. Our main findings can be summarised as follows.

	\begin{enumerate}[leftmargin=*]
		\item Our ALMA observations appear to trace the presence of nuclear starbursts, but are barely sensitive to extended spiral structure.  For all three galaxies, the CO emission appears $\gtrsim 30\%$ more compact than the rest-frame optical emission. The 1.3\,mm emission also appears more compact, by at least a factor of two, than the rest-frame optical emission of ALPS.1 and 3. In contrast, the 1.3\,mm emission of ALPS.2 appears as extended as the rest-frame optical, albeit at poor S/N.  The compact CO versus rest-frame optical emission implies the presence of a central, dense ISM component. Similarly, the compact 1.3\,mm emission of ALPS.1 is consistent with what would be observed for an extended disk galaxy hosting a nuclear starburst, tracing a high central gas column density (and potentially also a high central star formation rate). For both the CO and 1.3\,mm observations, temperature gradients would only serve to further to increase the contrast between the nuclear region and the more extended disks. Compared to ALPS.1, the more extended 1.3\,mm (versus 1.6\, \textmu m) emission of ALPS.2 may imply the existence of denser spiral structure at large radii. However, higher S/N data is needed to confirm this result. 
	 
		\item  Our observations of ALPS.3 imply that there may exist a population of sources where the vast majority of the gas and dust is within a compact, nuclear starburst region. Both the CO and dust continuum observations are significantly more compact than the rest-frame optical emission with no structure recovered at $\geq 3\sigma$ beyond $\sim 2$ kpc. For observations at the sensitivity and resolution of the 1.3\,mm data for ALPS.3 we would expect to be able to observe dense spiral structure with a column density of $N_\mathrm{H_2} \sim 10^{22}\, \mathrm{cm}^{-2}$ averaged over our synthesized beam, a value typical for extended local disk galaxies. 

		\item We investigate the dynamical properties of the three galaxies, based on our resolved observations of the CO(2-1) and CO(3-2) line emission. Our analysis indicates that the gas motion of all three galaxies is rotation-dominated. Based on the kinematic properties, we infer the dynamical masses and dark matter fractions within the region traced by the CO emission. Assuming CO-to-molecular gas conversion factors consistent with that of the Milky Way, we infer dark matter fractions of up to 0.6, consistent with simulations and observations of local galaxies with the same stellar masses.

	\end{enumerate}

	We conclude with a note of caution. Based on our analysis, we find that measurements of the dust and CO half-light radii are still hampered by low sensitivity observations. We barely recover the extended cool dust component for one source and only observe the CO emission out to $\sim 10$ kpc, despite substantial time investments with ALMA. At $3\sigma$ our data are sensitive to a molecular gas column density of $10^{22}\, \mathrm{cm}^{-2}$ per beam. Thus, we miss regions of gas that are less dense than this and/or smaller than the synthesized beam. To triple the sensitivity of the observations would require a tenfold increase in the ALMA time per source, for both the CO and continuum emission. Thus, significant additional investments with ALMA are required to observe the extended gas and dust disks of high-redshift galaxies. \\


\acknowledgments

We thank the anonymous referee for their prompt and constructive feedback. We thank Wiphu Rujopakarn for sharing the VLA data presented in Figure \ref{fig:thumbnails}. Thanks to Andrew Battisti for helpful discussions on the effects of dust attenuation on SED modeling and for performing the SED modeling with his adapted version of {\sc{MAGPHYS}}. We also thank Rachel Cochrane, Arjen van der Wel, Matus Rybak and Enrico di Teodoro for useful discussions and advice. We thank the developers of the fully-documented  publicly-available codes used here; Chien Peng (GALFIT), Ivan Mart\'{i}-Vidal (uvmultifit) and Enrico di Teodoro (3D Barolo). 

M.K. acknowledges support from the International Max Planck Research School for Astronomy and Cosmic Physics at Heidelberg University (IMPRS-HD).
Parts of this research were conducted by the Australian Research Council Centre of Excellence for All Sky Astrophysics in 3 Dimensions (ASTRO
3D), through project number CE170100013. F.W, Ml.N., and Ma.N. acknowledge support from the ERC Advanced Grant Cosmic Gas (740246). 

I.R.S acknowledges support from STFC (ST/T000244/1). E.d.C. gratefully acknowledges the Australian Research Council as the recipient of a Future Fellowship (project FT150100079). T.D-S acknowledges support from the CASSACA and CONICYT fund CAS-CONICYT Call 2018. Este trabajo cont \'o con el apoyo de CONICYT + PCI + INSTITUTO MAX PLANCK DE ASTRONOMIA MPG190030. D.R. acknowledges support from the National Science Foundation under grant numbers AST-1614213 and AST-1910107 and from the Alexander von Humboldt Foundation through a Humboldt Research Fellowship for Experienced Researchers. H.I. acknowledges support from JSPS KAKENHI Grant Number JP19K23462.

This paper makes use of the following ALMA data: 2012.1.00173.S, 2016.1.00324, 2016.1.00324.L \\ and 2017.1.00270.S. ALMA is a partnership of ESO (representing its member states), NSF (USA), and NINS (Japan), together with NRC (Canada), NSC and ASIAA (Taiwan), and KASI (Republic of Korea), in cooperation with the Republic of Chile. The Joint ALMA Observatory is operated by ESO, AUI/NRAO, and NAOJ.

\vspace{5mm}
\facilities{\ac{alma}, \ac{hst}}

\software{Astropy \citep{2013A&A...558A..33A},
Matplotlib \citep{10.1109/MCSE.2007.55},
CASA \citep{2010AAS...21547904R},
Topcat \citep{2005ASPC..347...29T},
GILDAS \citep{2013ascl.soft05010G}
}

\bibliographystyle{apj}
\bibliography{mybib}

\newpage

\appendix

\section{Additional target} 
	\label{sec:additional_target}

\begin{figure*}[h!]
		\centering
		\includegraphics[width=\textwidth, trim={1.5cm 2.cm 1.5cm 0cm},clip]{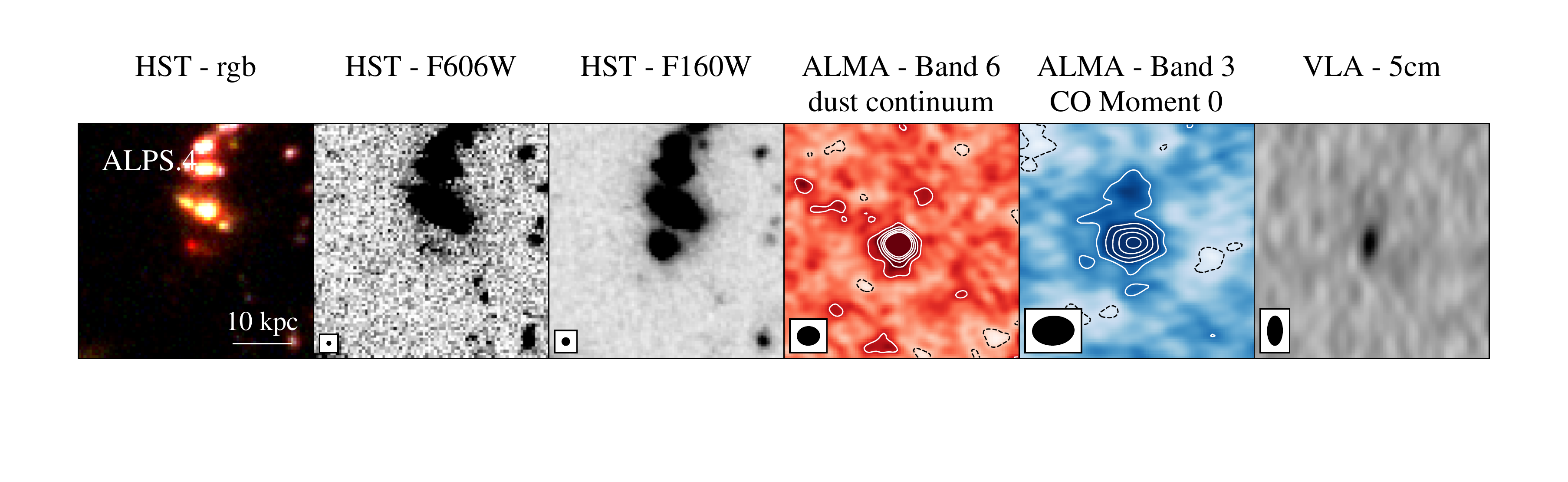}
		\caption{Multiwavelength data for the additional source, ASPECS LP 3mm.09, observed with ALPS.3. The dust-continuum and CO emission are unresolved and the source appears compact in the rest-frame optical (columns 1-3). Each panel depicts a $5" \times 5"$ region centred on the \ac{hst}-defined centre of the source. Columns, from left to right: \ac{hst} 435/775/105 color composite, \ac{hst}/F606W, \ac{hst}/F160W, Band 6 (1.3\,mm) dust continuum (combined \ac{aspecs} and \cite{2017MNRAS.466..861D} data), CO moment 0 map (from the combined ASPECS and ALPS data) and VLA - 5\,cm continuum flux. The contours for the Band 6 dust continuum and CO moment 0 start at $\pm 2\sigma$ and change in steps of $2\sigma$. Dashed black (solid white) lines show negative (positive) contours.  \label{fig:thumbnails_alps4}}
	\end{figure*}

	Along with ALPS.3, we observe an additional galaxy at an angular separation of $7\as56$. The additional source is labeled 3mm.09, in the 3\,mm  ASPECS Large Programme \cite{2019ApJ...882..139G} and UDF1 in the the 1.3\,mm programme of \cite{2017MNRAS.466..861D}. The galaxy is classified as an AGN, based on the X-Ray observations of \cite{2017ApJS..228....2L} and has a molecular gas to stellar mass fraction around unity \citep{2019ApJ...882..140B}. We show the \ac{hst} data along with our ALMA observations and the 5\,cm continuum in Figure \ref{fig:thumbnails_alps4}. Resolved observations of this source, at 870\,\textmu m, are presented and discussed in \cite{2019ApJ...882..107R}. They also model the \ac{hst} H160 emission, which they find is best-fit by the combination of a bright, point source and fainter S\'{e}rsic component with a half-light radius of $3.16\pm 0.17$\,kpc (potentially representing the AGN and disk, respectively). The centroid of the 870\,\textmu m emission \citep{2019ApJ...882..107R} and the 1.3\,mm data presented here, is co-located with the AGN/point source. We find no evidence for resolved CO or 1.3\,mm continuum emission for this source, nor do we find any evidence for rotation based on the channel maps or position-velocity diagrams. 

	The CO data for this source shows both the compact region observed in the 1.3\,mm dust continuum as well as an additional CO emitting region to the north, detected at $\gtrsim 3\sigma$. The northern region may be an additional CO emitter, which matches the position of the MUSE absorption line at the same redshift \citep[presented and discussed in Figure 2 and Appendix A of][]{2019ApJ...882..140B}. This additional or adjoined source exhibits a red core in the \ac{hst} images but is barely apparent in the rest-frame UV images and is not observed in the 870\,\textmu m data of \cite{2019ApJ...882..107R}. 


\section{SED Analysis} 
	\label{sec:sed_analysis}

	We briefly elaborate on some of the details of the SED modeling for our sources here. 


	To model the SEDs, we rely on the accuracy of the measured photometry. Two catalogues of photometry are available for the HUDF, spanning from the UV to the mid-IR (\emph{Spitzer} IRAC), those of \cite{2013ApJS..207...24G} and the 3D-\ac{hst} catalogue described in \cite{2014ApJS..214...24S}. The \ac{hst} fluxes from the two HUDF catalogues vary significantly across different bands for our sources. Although both \cite{2013ApJS..207...24G} and \cite{2014ApJS..214...24S} extract fluxes from \ac{hst} images matched to the source-detection image (the \ac{hst} F160W Band for our sources), their chosen aperture sizes differ. \cite{2014ApJS..214...24S} extract the source flux from circular apertures of 0."7 diameter 
	whereas \cite{2013ApJS..207...24G} infer the total flux from the isophotal area (i.e. the detection footprint in the F160W image), 
	The sources investigated here exhibit extended rest-frame optical emission, with F160W half-light radii of $0\as6-0\as98$ (see Table~\ref{tab:derived_prop}). Thus, the 3D-\ac{hst} measurements underestimate the flux of our sources, especially at shorter wavelengths (in the UV). We therefore choose to use the \ac{hst} and \emph{Spitzer}/IRAC photometry from the \cite{2013ApJS..207...24G} catalogue, which are consistent with the fluxes inferred from the XDF images using circular apertures enclosing our sources (cyan shaded regions and circles in Figure~\ref{fig:ex_sed}). For ALPS.1 and 2 the stellar masses inferred based on the \cite{2013ApJS..207...24G} catalogue are factor of 1.5 times lower than inferred from the \cite{2014ApJS..214...24S} catalogue whereas the SFRs are 1.5 times greater. Conversely, for ALPS.3 the \cite{2013ApJS..207...24G} photometry results in a stellar mass 1.5 times greater and SFR 1.5 times smaller than when using the \cite{2014ApJS..214...24S} photometry. 

	To model the SEDs use the {\sc{MAGPHYS}} algorithm. {{\sc{MAGPHYS}}} simulates the stellar emission of a galaxy using population sythesis models, based on a variety of star formation histories, and links the stellar energy absorbed and scattered by dust grains to the energy of the thermal emission. As discussed in \cite{2019A&A...632A..79B} and \cite{2019ApJ...882...61B},  the assumptions used to link the stellar and dust emission, such as a simple star formation history (SFH) and dust attenuation recipe, can significantly bias the derived parameters. Attenuation curves vary between galaxies \citep[e.g.][]{2016ApJ...827...20S,2018ApJ...859...11S,2018A&A...619A.135B} and flatten when the amount of obscuration or SFR increase \citep[e.g.][]{2010MNRAS.403...17J,2013MNRAS.432.2061C,2019ApJ...881...18R}.  

	In this work we have adopted two, adapted versions of the {\sc{MAGPHYS}} algorithm as the fits using the standard high-redshift {\sc{MAGPHYS}} algorithm left significant residuals. We apply the version of \cite{2019ApJ...882...61B} to ALPS.1 and 3. This adapted version includes a star formation history that both rises linearly at early ages and then declines exponentially (where before it was an exponentially declining function), has broader priors on the range of optical dust depths and equilibrium dust temperatures (to reflect high-redshift observations, particularly of sub-mm galaxies),  includes a prescription for absorption by the intergalactic medium and includes an additional component in the attenuation curve for the diffuse \ac{ism} to characterize the attenuation due to the 2175 \AA\ feature.

	We test the extent to which the inferred stellar masses and SFRs depend on different photometric bands. Based on the omission of different bands, we find that the peak of the near-infrared emission is particularly important in the stellar mass fits, i.e. the SED fits are highly sensitive to the IRAC photometry, particularly for ALPS.3. For ALPS.1, 2 and 3, respectively, omitting the IRAC photometry resulted in an inferred median stellar masses of $\sim1.5$, $\sim3$ and $\sim5$ times smaller than the median values inferred when using the full set of photometry.

	Multiple assumptions systematically bias the inferred stellar masses and SFRs, including the choice of aperture size for the \ac{hst} photometry (which change the inferred stellar masses by a factor of 1.5 - 2), our reliance on the accuracy of photometry at the peak of the stellar continuum emission (which change the inferred stellar masses by a factor of 1.5 - 5), the inclusion of an unconstrained AGN component (which change the inferred stellar masses by a factor of $\sim 2$ for $\xi_\mathrm{AGN}\sim 0.15$) and the applied dust attenuation curve (which also changes the inferred stellar masses by factor of $\sim 1.5-2$).  Based on these systematics we adopt an uncertainty floor of $\pm 0.3$ dex on the stellar masses and SFRs inferred here.


\section{Local Starbursts} 
	\label{sec:local_starbursts}

\begin{figure*}
		\centering
		\includegraphics[width=\textwidth, trim={3cm 1cm 3cm 1cm},clip]{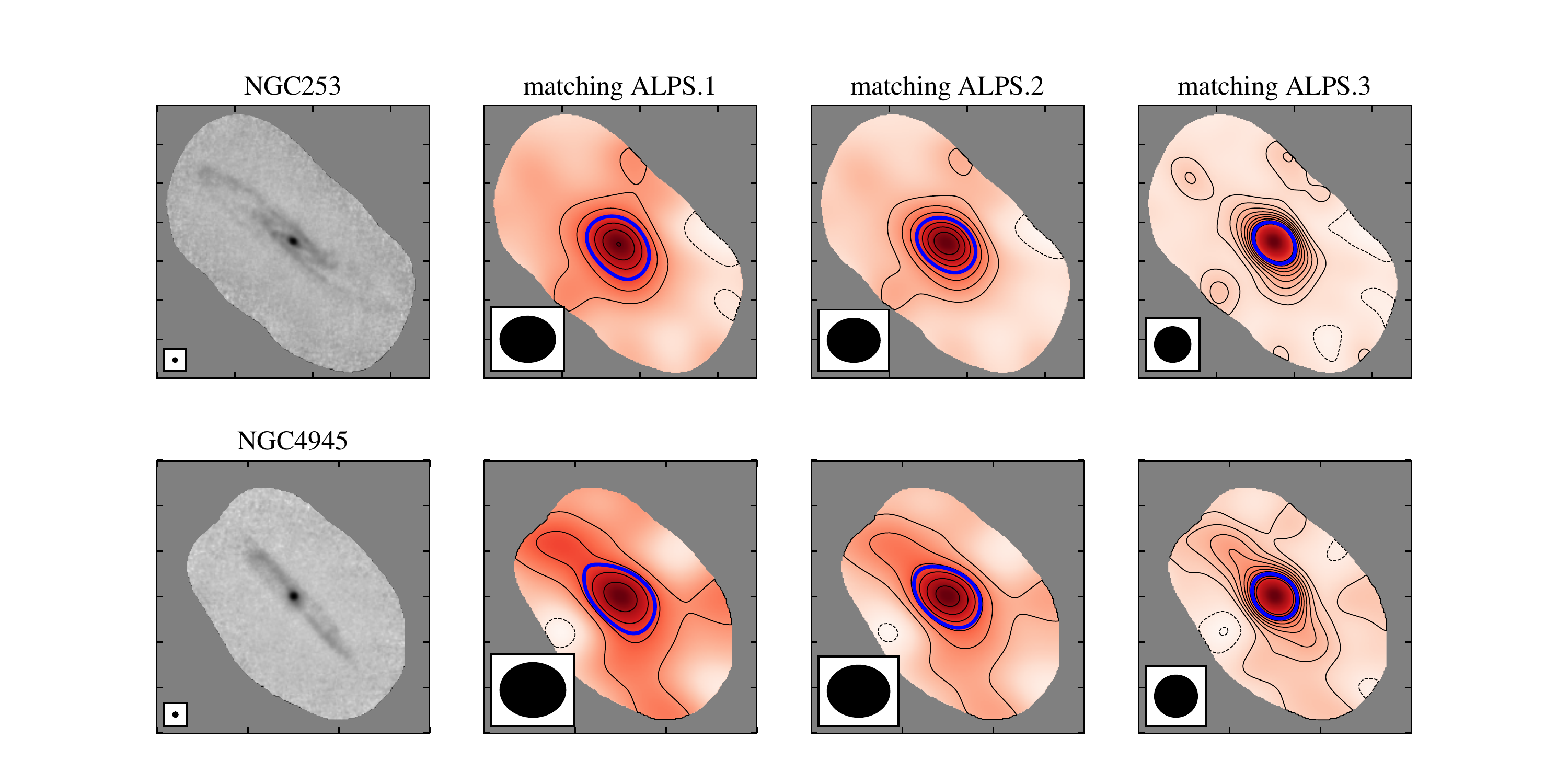}
		\caption{Mock observations of the local active galaxies, NGC 253 and NGC 4959, matching the total dust mass (i.e. measured flux) to the sources observed here. Left column: original Laboca observations. Second to fourth column: mock 1.3\,mm observations matching the rms and total flux density of ALPS.1, 2 and 3 respectively. Contours start at $\pm 2\sigma$ and go in steps of $2\sigma$. Solid (dashed) lines show positive (negative) contours. The half-peak emission contours are shown in blue. \label{fig:local_starbursts}}
	\end{figure*}

	To place the measured sizes for our data in context, we create mock 1.3\,mm maps of two local galaxies that harbour nuclear starbursts, NGC 253 and NGC 4945. We use the 870\,\textmu m LABOCA maps from \cite{2008A&A...490...77W}, smooth these to the resolution of the Band 6 data, scale the flux to match the total flux density of each of our sources and add the same level of noise as for our data (smoothed to the beam). We thereby assume that the distribution of the rest-frame 870 \textmu m matches that of the rest-frame 540, 509 and 350 \textmu m emission, respectively, for ALPS.1, 2 and 3. The results are shown in Figure \ref{fig:local_starbursts}. The mock observations are barely resolved, with the observed size at half the peak surface brightness (blue contours) matching the 1.3\,mm beam size. Only a portion of the brightest spiral arms appear to be recovered at $>3\sigma$.

\end{document}